\documentstyle[preprint,aps,prc]{revtex}

\newcounter{ref1}
\newcounter{ref2}
\begin{document}

\draft

\title{Microscopic analysis of quadrupole collective motion 
in Cr--
Fe nuclei\\
\Roman{ref1}. Renormalization of collective states 
and interacting-boson-model parameters}

\author{Hitoshi Nakada}
\address{Department of Physics, Faculty of Science, Chiba University\\
Yayoi-cho 1-33, Inage-ku, Chiba 263, Japan}
\author{Takaharu Otsuka}
\address{Department of Physics, Faculty of Science, 
University of Tokyo\\
Hongo 7-3-1, Bunkyo-ku, Tokyo 113, Japan}

\date{\today}

\maketitle

\begin{abstract}
We present a new method by which wavefunctions with simple structure
are renormalized so as to contain more complicated structure.
This method, called $H^n$--cooling method,
is applied to the study of the quadrupole collective motion
of $^{56}$Fe, $^{54}$Cr, $^{58}$Fe and $^{56}$Cr.
The shell-model wavefunctions of lowest-lying states of these nuclei
are well treated by this method.
By using the wavefunctions obtained via the $H^n$--cooling method,
IBM--2 parameters are derived from a realistic shell-model hamiltonian
and transition operators.
The Majorana interaction becomes sizably repulsive,
primarily as an effect of the renormalization.
The bosonic $E2$ effective-charges are enhanced 
due to the renormalization,
while a quenching occurs in the $M1$ and $M3$ parameters 
for proton bosons.
It is shown that the $\chi$ parameters take similar values
in the hamiltonian and in the $E2$ operator.
\end{abstract}


\section{Introduction}
\label{sec:intro}

Middle {\em pf}--shell nuclei provide us
with a precious testing ground
to understand various aspects of the quadrupole collective motion
from microscopic standpoints.
Computational difficulties in a realistic shell model
rise rapidly in general,
as the mass number increases.
The growing computer power, however, enables us to carry out
realistic shell-model calculations
in the middle {\em pf}--shell region.
On the other side, the middle {\em pf}--shell nuclei
seem to gain significant quadrupole collectivity,
which is a global and dominating feature of heavier nuclei.

Recently Sebe and the authors have reported
one of the most successful shell-model results
for $N=28-30$ nuclei\cite{ref:NOS91,ref:Nak91,ref:NSO94}.
The Kuo--Brown interaction\cite{ref:KBpf},
which had been derived from a realistic $NN$ potential
through $G$--matrix,
has been employed in these calculations,
together with a large configuration space
including excitations from the $0f_{7/2}$ orbit.
To be more precise, considering the following configuration,
\begin{equation} (0f_{7/2})^{n_1-k}(0f_{5/2}1p_{3/2}1p_{1/2})^{n_2+k},
\label{eq:config} \end{equation}
where $n_1=(Z-20)+8$ and $n_2=N-28$
for the $20<Z\leq 28\leq N<40$ nuclei,
we have adopted a space
consisting of all the $k=0,1$ and $2$ configurations.
It has been confirmed\cite{ref:NSO94} that, for even-even nuclei,
the energy levels are reproduced remarkably well
for $E_x<4$MeV.

The presence of mixed-symmetry states with respect to
the proton and neutron collective degrees of freedom
has been predicted
by the proton-neutron interacting boson model (IBM--2)\cite{ref:OAI}.
It has been pointed out\cite{ref:sciss-IBM} that
a mixed-symmetry $2^+$ state may lie lower
than the other mixed-symmetry states in spherical nuclei,
although the origin of such a low-lying mixed-symmetry $2^+$ level
has remained open.
Experimental studies have suggested
that the mixed-symmetry $2^+$ state exists
around $E_x=3$MeV in the Cr--Fe region\cite{ref:ms2-exp}.
A realistic shell-model analysis has been applied
to pin down the mixed-symmetry $2^+$ state 
of $^{56}$Fe\cite{ref:NOS91},
clarifying which states share substantial fractions
of the mixed-symmetry component.
It is of special interest
to study this type of collective modes more extensively,
on the basis of a realistic shell-model calculation.

The realistic shell-model hamiltonian couples
collective degrees of freedom to non-collective ones, in general.
When large-scale shell-model results are interpreted 
in terms of IBM--2,
it is important to incorporate,
through a certain renormalization procedure,
effects of relevant non-collective degrees of freedom
into the calculations made in the collective subspace.
This is an example of the general problem as to
how a complicated system can be described 
with a limited number of degrees of freedom
by taking into account a variety of correlations 
in an effective manner.
Rayleigh--Schr\"{o}dinger's perturbation theory 
constitutes a possible way,
by which the model wavefunction is modified.
The second-order perturbation has been applied
to renormalize the IBM--2 parameters\cite{ref:DPBD,ref:Miz92}.
Another way is Bloch--Horowitz's renormalization of operators,
in which operators, rather than wavefunctions,
are modified perturbatively
so as to carry the relevant correlation effects.
Some useful general theories have been developed
by extending Bloch--Horowitz's method:
Feshbach's projection method\cite{ref:Fes62} and
the folded-diagram theory\cite{ref:KO90}, for example.
These methods are, however, more or less 
based on the perturbation theory.
In the cases to be considered in this paper,
perturbative ways are inappropriate,
as is argued just below.

Our present goal is
an investigation of the quadrupole collective states
which are to be described within IBM--2,
in connection with the realistic shell model.
In the first approximation, the $s$-- and $d$--bosons in IBM--2
correspond to the collective $0^+$ ($S$) and $2^+$ ($D$) pairs
of valence like-nucleons\cite{ref:OAI}.
In Cr--Fe nuclei,
as is assumed in Ref.\cite{ref:ms2-exp},
the $S$-- and $D$--pairs normally comprise
only the $k=0$ configuration of Eq.(\ref{eq:config}),
because of the $Z=N=28$ magic number.
The realistic shell-model wavefunctions, however,
contain other configurations.
According to the realistic shell-model results,
the leakage out of the $k=0$ space is so significant
that even the $0^+_1$ and $2_1^+$ wavefunctions are not 
well enough covered
with this usual $SD$ space  ($<60$\%)\cite{ref:NSO94}.
In order that the $0^+_1$ and $2_1^+$ states
can be described within the IBM--2,
the correlations beyond the $SD$--pairs must be taken into account.
This relatively large $k>0$ fraction prevents perturbative ways
from being applicable.
A method beyond the perturbation theory is required.
It is commented that this situation takes place
because the $^{56}$Ni core is not very stiff.
Perturbative approaches may be legitimate in other mass-regions
in connecting IBM--2 to realistic shell-model.

We recall here that,
as far as several lowest-lying levels are concerned,
they are successfully reproduced
by the Horie--Ogawa hamiltonian\cite{ref:HO}
with only the $k=0$ configuration in the Cr--Fe region,
apart from the precise description
of the mixed-symmetry states\cite{ref:NOS91}.
Moreover, this $k=0$ shell-model result is connected 
with IBM--2 fairly well,
at least for $0^+_1$ and $2_1^+$\cite{ref:Halse}.
This fact suggests that,
even in more realistic cases with $k>0$ configurations,
the lowest-lying states may be described within IBM--2
through a proper renormalization.
In Ref.\cite{ref:NOS91} modified $SD$--pairs have been introduced,
and the fragmentation of the mixed-symmetry $2^+$ components
was clarified for $^{56}$Fe.

We introduce a new and yet simple method in this article.
It is applicable even to some cases
where the perturbation does not work well.
The method is applied to quadrupole collective states 
of Cr--Fe nuclei,
and the wavefunctions of those states are renormalized.
In addition to $^{56}$Fe and $^{54}$Cr,
on which the shell-model results have already been reported
in Ref.\cite{ref:NSO94},
$^{58}$Fe and $^{56}$Cr are studied.
Furthermore, by extending the OAI mapping\cite{ref:OAI},
the IBM--2 hamiltonian is derived 
from the realistic shell-model hamiltonian.
This is the first work of this sort,
while there have been many works evaluating IBM--2 parameters
from more schematic interactions,
for instance the surface-delta interaction.
The IBM--2 transition operators are obtained as well.
Renormalization effects on various IBM--2 parameters are discussed.
Focusing on the IBM--2 results more concisely,
we shall investigate properties of the mixed-symmetry states
in Cr--Fe region in the following paper\cite{ref:no.2}.

\section{$H$\lowercase{$^n$}--cooling method (H\lowercase{$^n$}CM)}
\label{sec:physics}

The present renormalization method is introduced in a general form,
in this section.
Some details of the procedure will be illustrated
in Section~\ref{sec:illust}.
Although this method may be applicable to other many-body problems,
we shall apply it, in this paper,
to elicit a collective space out of the shell-model space.
This collective space should correspond to that of IBM--2.

In $^{56}$Fe, for example,
we have a pair of proton holes and a pair of valence neutrons
within the description assuming the $^{56}$Ni core.
The $S$-- and $D$--pairs of protons are defined
as the $0^+$ and $2^+$ states of the $(0f_{7/2})^{-2}$ configuration,
while those of neutrons are collective $0^+$ and $2^+$ states
of the $(0f_{5/2}1p_{3/2}1p_{1/2})^2$ configuration.
Once the structure of the neutron $S$-- and $D$--pairs is given,
the $SD$ space is constructed
by these proton and neutron $S$-- and $D$--pairs.

We first introduce a subspace of the original Hilbert space.
This subspace is denoted by $W^{(0)}$.
In the application discussed in Section~\ref{sec:result},
the original space corresponds to the shell-model space,
and $W^{(0)}$ to the $SD$ space.
The bases belonging to $W^{(0)}$ are hereafter called 
{\it primary} bases,
while those outside $W^{(0)}$ {\it non-primary} bases.
It is required that the primary bases include basic dynamics already.
If there is a conserved quantum number $J$,
the space $W^{(0)}$ can be decomposed as
\begin{equation} W^{(0)} = \bigoplus_J W_J^{(0)}.
\end{equation}
In the practical case, $J$ represents nuclear spin.
In the following procedure, $W_J^{(0)}$'s with different $J$'s
never mix with one another,
reflecting the conservation law.

We consider a primary basis $\Psi_\lambda^{(0)} \in W_J^{(0)}$,
where $\lambda$ is the index of basis-state vectors for a given $J$.
The quantum number $J$ is not explicitly shown 
in $\Psi_\lambda^{(0)}$ for brevity.
The state $\Psi_\lambda^{(0)}$ evolves with time $t$
as $e^{-iHt}\left|\Psi_\lambda^{(0)}\right\rangle$,
where $H$ means the original hamiltonian.
When the inverse temperature $\beta=it$ with imaginary $t$ 
is employed,
the time evolution $e^{-iHt}$ is converted to cooling $e^{-\beta H}$.
In order to simplify the following discussion,
we assume without loss of generality
that all the eigenenergies are non-negative.
This situation is attained, if necessary,
by shifting the origin of energy.
The expectation value $\left\langle\Psi_\lambda^{(0)}\right|
e^{-\beta H}\left|\Psi_\lambda^{(0)}\right\rangle$,
which is a function of $\beta$,
is a superposition of exponentially-decreasing components
corresponding to eigenvalues of $H$.
The number of these components is much larger, in general,
than the dimension of $W^{(0)}$,
because a sizable fraction
of $e^{-\beta H}\left|\Psi_\lambda^{(0)}\right\rangle$
escapes out of $W^{(0)}$ with increasing $\beta$;
$e^{-\beta H}~W_J^{(0)} \ne W_J^{(0)}$
and therefore $e^{-\beta H}~W^{(0)} \ne W^{(0)}$.
We consider, in this paper, the situations in which
the primary bases form a major part of some low-lying states
and then $\left\langle\Psi_\lambda^{(0)}\right| e^{-\beta H} 
\left|\Psi_\lambda^{(0)}\right\rangle$
is dominated by one or a few slowly-decreasing components,
while fast-decaying components are superposed
with far smaller amplitudes.
If we choose appropriate states from non-primary bases,
only a small number of them will have a sizable 
mixing with $\Psi_\lambda^{(0)}$.
These non-primary but relevant bases are here expressed
as $\phi_\lambda^{(\nu)}$.
The superscript $(\nu)$ is used
as an index of degree of the coupling to $\Psi_\lambda^{(0)}$,
whose meaning will be specified later in this section.
By taking into account the influence of the $\phi$--bases,
the wavefunction of $\Psi_\lambda^{(0)}$ will be renormalized as
\begin{equation} \Psi_\lambda \propto \Psi_\lambda^{(0)}
+ \sum_\nu c_{\nu,\lambda} \phi_\lambda^{(\nu)} ,
\label{eq:wf-ren} \end{equation}
where $c_{\nu,\lambda}$ represents mixing amplitude
of the $\phi$--basis.
The basis $\Psi_\lambda$ is constructed
so as to contain higher-energy components
with significantly small amplitudes.
By doing this, rapidly-decreasing components
in $\left\langle\Psi_\lambda\right| e^{-\beta H} 
\left|\Psi_\lambda\right\rangle$ can be made negligibly small.
Then the exact low-lying eigenstates will be reproduced to a good
approximation 
by appropriate linear combinations of $\Psi_\lambda$'s.
The following discussion will exhibit 
how to choose $\phi_\lambda^{(\nu)}$ efficiently
and how to evaluate $c_{\nu,\lambda}$.
With an adequate set of $\phi_\lambda^{(\nu)}$'s
and $c_{\nu,\lambda}$'s,
the truncation up to relatively small $\nu$ in Eq.(\ref{eq:wf-ren})
is expected to yield good renormalized bases,
as will be shown with concrete examples.

In the present method, it is required that the Hilbert space
$W_J\equiv\{\Psi_\lambda;\lambda=1,2,\cdots\}$,
which consists of the renormalized bases,
fulfills the approximate relation
\begin{equation} e^{-\beta H}~W_J \approx W_J,
\label{eq:Vclose} \end{equation}
up to a reasonably large $\beta$.
By defining the total space of the renormalized bases by
\begin{equation} W \equiv \bigoplus_J W_J, \end{equation}
Eq.(\ref{eq:Vclose}) can be expressed as
\begin{equation} e^{-\beta H}~W \approx W.
\label{eq:Vclose0} \end{equation}
We look for a $W$ which satisfies Eq.(\ref{eq:Vclose0})
and, at the same time,
remains a rather small subspace of the full Hilbert space.
Eq.(\ref{eq:Vclose}) or (\ref{eq:Vclose0}) indicates
approximate closure of the renormalized space.
The closure is exactly satisfied
if $W$ consists of eigenstates of $H$.
Such a construction, however, only means a calculation 
in the full space.
We are here seeking to obtain $W_J$
with a limited number of the $\phi$--bases in Eq.(\ref{eq:wf-ren}),
discarding degrees of freedom coupled to the primary bases 
only weakly.
For this purpose,
we shall trace how the original basis $\Psi_\lambda^{(0)}$
evolves by $e^{-\beta H}$.

We consider a small $\beta$ by rewriting $\beta$ as $\Delta\beta$
for the time being,
although it is not essential as discussed later.
By expanding $e^{-\Delta\beta H}$ into the power series 
of $\Delta\beta$,
the cooling of $\Psi_\lambda^{(0)}$ gives
\begin{equation}
 e^{-\Delta\beta H}\left|\Psi_\lambda^{(0)}\right\rangle
 = \left[\sum_{\nu=0}^n {{(-\Delta\beta)^\nu}\over{\nu!}} H^\nu
 + {\hat O}\left((\Delta\beta)^{n+1}\right)\right]
 \left|\Psi_\lambda^{(0)}\right\rangle ,
\label{eq:short} \end{equation}
where ${\hat O}\left((\Delta\beta)^\nu\right)$ represents
an operator with the order of $(\Delta\beta)^\nu$.
We define the $\phi_\lambda^{(\nu)}$ bases (see Eq.(\ref{eq:wf-ren}))
from the right-hand side,
\begin{equation} \left|\phi_\lambda^{(\nu)}\right\rangle \equiv
P^{\cal O}\cdot H^\nu\left|\Psi_\lambda^{(0)}\right\rangle .
\label{eq:phi} \end{equation}
Here $P^{\cal O}$ stands for an appropriate orthonormalization,
whose concrete definition will be given in Section~\ref{sec:illust}.
The basis $\phi_\lambda^{(1)}$ directly couples
to $\Psi_\lambda^{(0)}$ via $H$,
exhausting the coupling leading out of $W_J^{(0)}$.
The next basis $\phi_\lambda^{(2)}$ affects the primary state
via its coupling to $\phi_\lambda^{(1)}$.
In this manner, important bases are extracted one after another.
Note that all the $\phi$--bases carry the same quantum number $J$
as $\Psi_\lambda^{(0)}$.

Since the shell model is defined 
as a finite dimensional many-body problem,
the cooling operator $e^{-\Delta\beta H}$ does not need 
infinite series expansion.
Moreover, since we have postulated
that the primary bases include basic dynamics,
the number of relevant degrees of freedom
which couple to the primary bases can be relatively small.
Making good use of these features, 
$e^{-\Delta\beta H}$ is handled by the power-series expansion
as in Eq.(\ref{eq:short}),
and the $\phi$--bases are generated by Eq.(\ref{eq:phi}).

It has been known that the Lanczos diagonalization algorithm 
is efficient
to obtain eigenenergies and eigenfunctions of low-lying states.
We here try to utilize the advantage of the Lanczos method.
The Lanczos method can be derived via the power-series expansion
of $e^{-\Delta\beta H}$ acting on an arbitrary basis.
Besides a difference in the $P^{\cal O}$ operator
(see Section~\ref{sec:illust}),
the $\phi$--bases in Eq.(\ref{eq:phi}) have basically the same form
as the Lanczos bases generated from $\Psi_\lambda^{(0)}$.
However, it is a key point of the present method
that the primary bases contain basic dynamics of the system.
In other words, the primary bases form
the main part of the wavefunctions of the low-lying states
under interest.
Thereby we can regard the current procedure as a renormalization.
The relation of the present method to the Lanczos method
will be discussed further in Section~\ref{sec:discuss}.

Let us begin with a simple case in which
there is just a single state $\Psi^{(0)}$ in $W_J^{(0)}$.
We do not need the label $\lambda$ in this case.
The $P^{\cal O}$ operator in Eq.(\ref{eq:phi})
expresses the Gram--Schmidt orthogonalization
to $\Psi^{(0)}$ and $\phi^{(\nu')}$'s with $\nu'<\nu$.
We thus generate a subspace $\Gamma^{(n)} \equiv \left\{ \Psi^{(0)},
\phi^{(1)}, \phi^{(2)}, \cdots, \phi^{(n)} \right\}$,
corresponding to the order of $(\Delta\beta)^n$.

A renormalized wavefunction (see Eq.(\ref{eq:wf-ren}))
is introduced within $\Gamma^{(n)}$,
\begin{equation} \Psi^{(n)} \propto \Psi^{(0)}
+ \sum_{\nu=1}^n c_\nu \phi^{(\nu)} ,
\label{eq:wf-ren1} \end{equation}
with the amplitude $c_\nu$'s to be determined.
The submatrix of $H$ for the subspace $\Gamma^{(n)}$ is constructed,
and the eigenvector associated with the lowest eigenvalue
is adopted as the renormalized basis $\Psi^{(n)}$
in Eq.(\ref{eq:wf-ren1}).
It is noticed that the mixing amplitude $c_\nu$ thus obtained
depends on $n$,
though this dependence is not explicitly shown here.
The basis $\Psi^{(n)}$ yields
\begin{equation} e^{-\Delta\beta H} \left|\Psi^{(n)}\right\rangle
   = e^{-\Delta\beta E^{(n)}} \left|\Psi^{(n)}\right\rangle +
 \left[{\hat O}\left((\Delta\beta)^{n+1}\right)\right]
  \left|\Psi^{(0)}\right\rangle ,
\label{eq:short-ren1} \end{equation}
where $E^{(n)}=\left\langle\Psi^{(n)}\right|H
\left|\Psi^{(n)}\right\rangle$.
By this procedure, rapidly-decreasing components
become substantially smaller
in $\left\langle\Psi^{(n)}\right|e^{-\Delta\beta H}
\left|\Psi^{(n)}\right\rangle$
than in $\left\langle\Psi^{(0)}\right|e^{-\Delta\beta H}
\left|\Psi^{(0)}\right\rangle$.
The $\Psi^{(0)}$ state is thus cooled down.
Increasing $n$ step by step, we can monitor
what components are adopted in higher-order steps.

When we have $l(>1)$ bases in $W_J^{(0)}$,
the renormalized basis $\Psi_\lambda^{(n)}$ is obtained
for each $\lambda$
by diagonalizing a submatrix of $H$ in $\Gamma_\lambda^{(n)}$,
which is generated from $\Psi_\lambda^{(0)}$.
The $\Psi_\lambda^{(n)}$ $(\lambda=1,\cdots,l)$ bases 
span a space $W_J^{(n)}$.
However, $H$ may produce cross-over couplings
among bases in $\Gamma_\lambda^{(n)}$ with different $\lambda$,
which gives rise to a non-orthogonality
between bases with different $\lambda$.
In order to avoid this, $\Gamma_\lambda^{(n)}$ is created
so as to be orthogonal to $\Gamma_{\lambda'}^{(n)}$
if $\lambda\ne\lambda'$,
by carrying out an orthonormalization.
This orthogonalization will be illustrated concretely
in Section~\ref{sec:illust}.
Although this modification can generally break
the relation (\ref{eq:short-ren1}) for individual basis,
the orthonormalization can be made
(see Section~\ref{sec:illust}) so that
a similar condition should be satisfied for the space $W_J^{(n)}$,
\begin{equation} e^{-\Delta\beta H}~W_J^{(n)}
 = W_J^{(n)} + \left[{\hat O}\left((\Delta\beta)^{n+1}\right)\right]
 ~W_J^{(0)} .
\label{eq:short-ren2} \end{equation}
This indicates that $W_J^{(n)}$ fulfills Eq.(\ref{eq:Vclose})
up to $O\left((\Delta\beta)^n\right)$.
The entire space of the renormalized bases at the order $n$
is then defined as
\begin{equation} W^{(n)} \equiv \bigoplus_J W_J^{(n)}.
\end{equation}

The cooling Eq.(\ref{eq:short-ren2}) is carried out step by step,
through the power-series expansion (\ref{eq:short}).
We shall call the present procedure $H^n$--cooling method (H$^n$CM).
The H$^n$CM gives a wavefunction renormalization,
incorporating dynamical correlations contained in $H$.
As far as the H$^1$CM is concerned,
the state $e^{-\beta H}\left|\Psi_\lambda^{(0)}\right\rangle$
with small $\beta$ is decomposed
in terms of the $n=1$ basis $\Psi_\lambda^{(1)}$ and the rest.
The latter has higher energy than the former,
giving rise to the faster-decreasing component.
Analogously, for a general $n$,
the H$^n$CM process produces $n$ faster-decreasing components
in addition to the slowest-decreasing component
({\it i.e.}, $\Psi_\lambda^{(n)}$). 
The closure of the renormalized subspace $W^{(n)}$
is fulfilled up to $O\left((\Delta\beta)^n\right)$,
as is shown in Eq.(\ref{eq:short-ren2}).
The larger $n$ assures the better approximation
from the viewpoint of the condition (\ref{eq:Vclose}) 
or (\ref{eq:Vclose0}).
If $W^{(n)}$ converges with $n$,
no new basis is created by $H$ acting on this subspace.
The convergence then becomes independent of $\beta$,
which means that Eq.(\ref{eq:Vclose0}) holds 
for a general value of $\beta$,
not only for $\Delta\beta$.

Although a cooling can be made only by operating $e^{-\Delta\beta H}$
on $\Psi_\lambda^{(0)}$, in the H$^n$CM the
diagonalization is performed within $\Gamma_\lambda^{(n)}$
for each step.
This accelerates the cooling to an appreciable extent,
since the diagonalization is equivalent to the full cooling
within the relevant subspace.
Moreover, as far as the dimension of $H$ is finite,
bases are exhausted at finite $n$.
Therefore, the $\beta\rightarrow\infty$ limit,
which is required for the full cooling in infinite-dimensional cases,
is not necessary.
Because of these properties,
all the major components for low-lying states are generated
with relatively small $\beta$,
and the H$^n$CM is expected to be efficient 
even with rather small $n$.
We shall see it in practice in Sections~\ref{sec:result}.

Here we should add the comment that some nuclear collective states,
for which the H$^n$CM will be used, are not necessarily
the lowest-lying state with a specific spin-parity.
In such cases the term `cooling' may not be appropriate,
and some caution will be necessary in applying the present method.
A prescription will be shown in Section~\ref{sec:discuss},
while the actual case of the Cr--Fe nuclei will be presented
in Ref.\cite{ref:no.2}.

\section{Illustration of H\lowercase{$^n$}CM}
\label{sec:illust}

The H$^n$CM is illustrated in some detail with an example:
let us consider the set spanned by the $SD$--states
with $J^P=2^+$ in $^{56}$Fe.

As has been mentioned in the preceding section,
the proton $S$-- and $D$--pairs have
the $(0f_{7/2})^{-2}$ configuration in $^{56}$Fe,
while the neutron pairs have
the $(0f_{5/2}1p_{3/2}1p_{1/2})^2$ configuration.
The set of $2^+$ states within this $SD$ space of $^{56}$Fe
comprises the following bases,
\begin{mathletters} \label{eq:coll-states}
\begin{equation} \left| 2^+(SD); F=1 \right\rangle =
 {1 \over {\sqrt 2}} \left( |D_\pi \rangle \otimes |S_\nu \rangle
  + |S_\pi \rangle \otimes |D_\nu \rangle \right) ,
\label{eq:coll-state-a} \end{equation}
\begin{equation} \left| 2^+(D^2); F=1 \right\rangle =
 \left[ |D_\pi \rangle \otimes |D_\nu \rangle \right]^{(2)} ,
\label{eq:coll-state-b} \end{equation}
\begin{equation} \left| 2^+(SD); F=0 \right\rangle =
 {1 \over {\sqrt 2}} \left( |D_\pi \rangle \otimes |S_\nu \rangle
  - |S_\pi \rangle \otimes |D_\nu \rangle \right) .
\label{eq:coll-state-c} \end{equation}
\end{mathletters}
These bases straightforwardly correspond to the IBM--2 bases
through the OAI mapping\cite{ref:OAI}.
In the following discussions,
the proton-neutron property of the wavefunctions
is taken into consideration,
so that the states should correspond to the IBM--2 states 
with good $F$--spin.
The variable $F$ on the left-hand side of Eq.(\ref{eq:coll-states})
indicates the $F$--spin value of the corresponding IBM basis
after the OAI-mapping\cite{ref:OAI}.
The maximum of $F$ is obtained by $F_{\rm max}={1\over 2}N^{\rm B}$,
where $N^{\rm B}=N_\pi^{\rm B}+N_\nu^{\rm B}$ is
the total number of $SD$--pairs,
which is the same as the total boson number in the IBM--2, 
for each nucleus.
The states with $F=F_{\rm max}$
are called totally-symmetric states in the IBM--2,
while those with $F=F_{\rm max}-1$ mixed-symmetry states.
We refer here the bases in the $SD$ fermion space 
in an analogous manner.
Note that $F_{\rm max}=1$ in $^{56}$Fe.
Though the above $\left| 2^+ (SD) ; F=0 \right\rangle$ state is
a totally anti-symmetric state,
it is called mixed-symmetry state in this article,
because it belongs to the class of the $F=F_{\rm max}-1$ states.

For each nucleus we shall consider the complete set 
of orthonormal bases belonging to $W_J^{(0)}$ ({\it i.e.}, 
the $SD$ space with a specific spin);
$\Psi_1^{(0)}$, $\Psi_2^{(0)}$, $\cdots$, $\Psi_l^{(0)}$.
In the above case of Eq.(\ref{eq:coll-states}),
the $\Psi^{(0)}$'s turn out to be
\begin{equation}
\left|\Psi_1^{(0)}\right\rangle = \left|2^+(SD);F=1\right\rangle,~~
\left|\Psi_2^{(0)}\right\rangle = \left|2^+(D^2);F=1\right\rangle,~~
\left|\Psi_3^{(0)}\right\rangle = \left|2^+(SD);F=0\right\rangle.
\label{eq:coll-states2} \end{equation}
The ordering of the bases may affect the process of the H$^n$CM,
as will become transparent below.
In this example, the $SD$--bases are ordered so
that the class of states with full $F$--spin symmetry 
($F=F_{\rm max}$) should come first,
those with next highest $F$--spin ($F=F_{\rm max}-1$) come second,
and so forth.
Within each sector of a given $F$--spin,
the bases are placed from the lower seniority to the higher,
similarly to the OAI mapping.

In the H$^n$CM, the $\phi$-bases of Eq.(\ref{eq:phi})
as well as the $\Psi^{(0)}$'s are generated in the following order,
\begin{equation} \begin{array}{cccc}
\Psi_1^{(0)},&\Psi_2^{(0)},&\cdots,&\Psi_l^{(0)},\\
\phi_1^{(1)},&\phi_2^{(1)},&\cdots,&\phi_l^{(1)},\\
\phi_1^{(2)},&\phi_2^{(2)},&\cdots,&\phi_l^{(2)},\\
\multicolumn{4}{c}{\cdots \cdots \cdots,}\\
\phi_1^{(n)},&\phi_2^{(n)},&\cdots,&\phi_l^{(n)}.
\end{array} \label{eq:ren-bases} \end{equation}
Recall that $\phi_\lambda^{(\nu)}$ is generated
from $H^\nu \Psi_\lambda^{(0)}$,
apart from the orthonormalization by $P^{\cal O}$.
We define the $P^{\cal O}$ operator in Eq.(\ref{eq:phi}) as follows.
The first $l$ bases are $\Psi_1^{(0)}$, $\Psi_2^{(0)}$, $\cdots$,
$\Psi_l^{(0)}$, which are already orthonormal,
and $P^{\cal O}$ acts as the unity for them.
The $(l+1)$-th basis is $\phi_1^{(1)}$ generated from $H\Psi_1^{(0)}$
with the Gram--Schmidt orthogonalization to $\Psi_1^{(0)}$,
$\Psi_2^{(0)}$, $\cdots$, $\Psi_l^{(0)}$.
Namely,
\begin{equation} \phi_1^{(1)} \equiv
P^{\cal O}_{\{\Psi_1^{(0)},\cdots,\Psi_l^{(0)}\}}
\cdot H\Psi_1^{(0)}, \label{eq:orth1} \end{equation}
where $P^{\cal O}_{\{~\}}$ represents orthogonalization
to the states specified in the curly bracket,
together with the normalization.
The $(l+2)$-th basis $\phi_2^{(1)}$ is created similarly,
except that it should be orthogonal also to $\phi_1^{(1)}$,
\begin{equation} \phi_2^{(1)} \equiv
P^{\cal O}_{\{\Psi_1^{(0)},\cdots,\Psi_l^{(0)}, \phi_1^{(1)}\}}
\cdot H\Psi_2^{(0)}. \label{eq:orth2} \end{equation}
One can repeat the procedure until all the orthonormal bases
in Eq.(\ref{eq:ren-bases}) are obtained.
Each $\phi$--basis can be represented explicitly as
\begin{mathletters} \label{eq:orth}
\begin{equation} \phi_\lambda^{(\nu)} \equiv
P^{\cal O}_{\{\Psi_1^{(0)},\cdots,\Psi_l^{(0)},
 \phi_1^{(1)},\cdots,\phi_l^{(\nu-1)}\}}
\cdot H^\nu \Psi_\lambda^{(0)}~~(\mbox{for $\lambda=1$}), 
\end{equation}
\begin{equation} \phi_\lambda^{(\nu)} \equiv
P^{\cal O}_{\{\Psi_1^{(0)},\cdots,\Psi_l^{(0)},
 \phi_1^{(1)},\cdots,\phi_l^{(\nu-1)},\phi_1^{\nu)},
 \cdots,\phi_{\lambda-1}^{(\nu)}\}}
\cdot H^\nu \Psi_\lambda^{(0)}~~(\mbox{for $\lambda\neq 1$}). 
\end{equation}
\end{mathletters}
Note that, as $n$ becomes larger, some bases may vanish
due to the orthogonalization.
The bases in Eq.(\ref{eq:ren-bases}) are produced in this manner,
by carrying out the Gram--Schmidt orthogonalization on them
successively.

We then consider a subset $\Gamma_\lambda^{(n)}
\equiv\left\{ \Psi_\lambda^{(0)},\phi_\lambda^{(1)},
\phi_\lambda^{(2)},\cdots,\phi_\lambda^{(n)} \right\}$
for each $\lambda(=1,2,\cdots,l)$.
It should be noticed that $\Gamma_\lambda^{(n)}$ is spanned 
by the bases
constituting the $\lambda$-th column of Eq.(\ref{eq:ren-bases}).
In order to obtain a renormalized basis $\Psi_\lambda^{(n)}$
({\it i.e.}, to evaluate $c_{\nu,\lambda}$ of Eq.(\ref{eq:wf-ren})),
we construct a submatrix of $H$
within this subspace $\Gamma_\lambda^{(n)}$.
After diagonalizing this submatrix, the lowest eigenstate
is taken as the $\lambda$-th basis in $W_J^{(n)}$.
In this manner, $\Psi_\lambda^{(1)}$ is obtained from
$\Gamma_\lambda^{(1)}=\left\{ \Psi_\lambda^{(0)},
\phi_\lambda^{(1)}\right\}$,
$\Psi_\lambda^{(2)}$ from $\Gamma_\lambda^{(2)}=
\left\{ \Psi_\lambda^{(0)},\phi_\lambda^{(1)},
\phi_\lambda^{(2)}\right\}$,
and so forth.
The H$^n$CM space $W^{(n)}$ is spanned by the bases thus obtained.

The H$^n$CM will be useful
for extracting some simple structural features
from complicated shell-model wavefunctions.
Since the renormalized wavefunctions of the $SD$ states
are explicitly constructed,
it is possible to compare them directly
to the shell-model wavefunctions.
It is also straightforward to evaluate matrix elements
of a given operator in the space $W^{(n)}$.
Although the larger $n$ implies the better closure of the subspace
from the viewpoint of Eq.(\ref{eq:Vclose}),
we consider $n\leq 2$ cases in the following application.

\section{Application of H\lowercase{$^n$}CM to $SD$ space
in C\lowercase{r}--F\lowercase{e} nuclei}
\label{sec:result}

The H$^n$CM is applied and tested numerically in Cr--Fe nuclei,
starting from the $SD$--pair states.

As has been shown in Ref.\cite{ref:NSO94},
the shell-model calculation with the Kuo--Brown realistic hamiltonian
in the $k\leq 2$ space
successfully reproduces the observed states up to $E_x\simeq 4$MeV
in $^{54}$Cr and $^{56}$Fe.
While the proton $S$-- and $D$--pairs are uniquely determined
by the $(0f_{7/2})^{-2}$ configuration,
the structure of the neutron pairs has to be fixed.
In $^{56}$Fe, the structure of $|S_\nu \rangle$ is determined
so as to maximize the overlap
between the $|S_\pi \rangle \otimes |S_\nu \rangle$ state
and the shell-model $0_1^+$ state.
The structure of $|D_\nu \rangle$ is determined
so that the overlap between $|S_\pi \rangle \otimes |D_\nu \rangle$
and the shell-model $2_1^+$ state should be maximum.
The neutron pair structure for $^{54}$Cr is fixed
so as for $|S_\pi^2 \rangle \otimes |S_\nu \rangle$
($|S_\pi^2 \rangle \otimes |D_\nu \rangle$) to have maximum overlap
with the shell-model $0_1^+$ ($2_1^+$) state.
The structure of neutron pairs is slightly different
between $^{56}$Fe and $^{54}$Cr.
Note that the seniority projection\cite{ref:OAI} is carried out
when $|D_\pi^2 ; J\rangle$ is produced in Cr.

The H$^n$CM is applied, starting with the collective space
composed of these $SD$--pairs.
The H$^n$CM is driven by the shell-model hamiltonian
in the $k\leq 2$ space.
The subspace $W_J^{(0)}$ is spanned
by the products of the above $SD$--pairs
with angular momentum $J$.
Any basis in $W^{(0)}$ carries the lowest isospin,
and the isospin is conserved during the H$^n$CM.
The primary bases $\Psi_\lambda^{(0)}$'s in $W_J^{(0)}$
are put in the order of $F$--spin and seniority,
as has been mentioned in the preceding section.

For $n\ge 1$, $\Psi_\lambda^{(n)}$'s imply renormalized $SD$ states,
while they are referred to by the corresponding $SD$--pair structure
of $\Psi_\lambda^{(0)}$.
Once $\Psi_\lambda^{(n)}$'s are obtained,
eigenstates within $W_J^{(n)}$ are calculated
by diagonalizing the submatrix of the hamiltonian
whose elements are $\left\langle\Psi_{\lambda'}^{(n)}\right|
H\left|\Psi_\lambda^{(n)}\right\rangle$.
In Fig.\ref{fig:Fe56ren}(a),
the energies of lowest $0^+$ and $2^+$ states 
within the subspace $W^{(n)}$ are shown,
in comparison with those of the shell model, for $^{56}$Fe.
The origin of energy is set to the shell-model ground-state energy.
The unrenormalized $SD$--states gives far higher energies
(the most leftward sector)
than the shell-model eigenenergies (the most rightward sector).
In this regard,
the bare $SD$--states ({\it i.e.}, the primary bases)
are insufficient.
The renormalization via the H$^n$CM reduces 
this discrepancy of energies quite efficiently.
The H$^2$CM appears to recover
the shell-model $0^+_1$ and $2^+_1$ states satisfactorily well.
This observation is confirmed by direct comparison 
of the wavefunctions.
Table~\ref{tab:ovl} shows overlaps between the $SD$--states
and the shell-model eigenstates for the lowest $0^+$ and $2^+$ states.
The unrenormalized states are certainly different
from the shell-model eigenstates.
A large part of the discrepancy comes from the $k>0$ configurations.
On the contrary, the wavefunctions after the H$^2$CM are quite close
to the corresponding shell-model ones,
having more than $90\%$ overlaps.

Figure~\ref{fig:Fe56ren}(b) depicts energy levels
for which energies are measured
from the ground state defined in each space.
It is remarked that the levels without the renormalization
resemble the ones after the H$^2$CM,
whereas the H$^1$CM spectrum is certainly different.
The $2^+_1$ excitation-energy in the original $SD$ space
is in good agreement with the shell-model result,
and therefore with experiments.
In the H$^1$CM result, the $0^+_1$ state is greatly lowered
owing to the coupling to non-$SD$ ({\it i.e.}, non-primary) degrees 
of freedom.
Though the $2^+_1$ state is also lowered,
the non-$SD$ effect is smaller than in $0^+_1$.
Some additional non-$SD$ effect is absorbed by the H$^2$CM,
which recovers the $2^+_1$ excitation energy.
In an analogous manner, 
as far as the excitation spectra are concerned,
the result after the H$^2$CM is close to the unrenormalized one
for most lower-lying states.
There is a certain difference in the $1^+$ state.
We shall return to this point later.
The $4_1^+$ state appears to be too high, even in the H$^2$CM result.
This state seems to be largely influenced 
by the $G$--pair degrees of freedom.
Although some parts of them are included 
in the renormalized wavefunctions,
they are not yet sufficient for compensating the whole influence 
of the $G$--pairs.

In Cr nuclei, there is no $|D_\pi^2;0^+\rangle$ basis
({\it i.e.}, $0^+$ state with $(0f_{7/2})^4$ and seniority 4).
It is not always possible, thereby,
to create $\Psi_\lambda^{(0)}$ basis having a good $F$--spin value.
For instance, the $|0^+(SD^2)\rangle$ bases with good $F$--spins are
\begin{mathletters}\label{eq:Cr54a}
\begin{equation} \left|0^+(SD^2); F=\case{3}{2}\right\rangle
= {1\over{\sqrt 3}} \left( |D_\pi^2; 0^+ \rangle \otimes
 |S_\nu \rangle
 + \sqrt 2 \left[|S_\pi D_\pi \rangle
  \otimes |D_\nu \rangle\right]^{(0)} \right) ,
\label{eq:Cr54a-a} \end{equation}
\begin{equation} \left|0^+(SD^2); F=\case{1}{2}\right\rangle
= {1\over{\sqrt 3}} \left( \sqrt 2
 |D_\pi^2; 0^+ \rangle \otimes |S_\nu \rangle
 - \left[|S_\pi D_\pi \rangle \otimes |D_\nu \rangle\right]^{(0)}
  \right) .
\label{eq:Cr54a-b} \end{equation}
\end{mathletters}
Since $|D_\pi^2;0^+ \rangle$ does not exist in the present case,
only a single $|0^+(SD^2)\rangle$ basis is possible,
and we introduce
\begin{equation} \left|0^+(SD^2); F=\case{3}{2}\right\rangle'
\equiv \left[|S_\pi D_\pi \rangle \otimes |D_\nu \rangle\right]^{(0)} .
\label{eq:Cr54b} \end{equation}
We replace the higher $F$--spin basis (\ref{eq:Cr54a-a})
by (\ref{eq:Cr54b}).
The lower $F$--spin basis is also subject to similar changes,
and should be modified with proper orthogonalization
to the basis assigned with higher $F$ value.
This orthogonalization, however,
annihilates the basis corresponding to Eq.(\ref{eq:Cr54a-b}),
because of the lack of the $|D_\pi^2;0^+ \rangle$ component.
The following primary bases are thus obtained,
\begin{equation}
\left|\Psi_1^{(0)}\right\rangle=\left|0^+(S^3);F={3\over 
2}\right\rangle,~~
\left|\Psi_2^{(0)}\right\rangle=\left|0^+(SD^2);F={3\over 
2}\right\rangle',~~
\left|\Psi_3^{(0)}\right\rangle=\left|0^+(D^3);F={3\over 2}
\right\rangle.
\label{eq:Cr54a2}\end{equation}
The $|2^+(D^3)\rangle$ bases are handled in an analogous manner.
The energy levels in the $SD$ space of $^{54}$Cr thus constructed
are shown in Fig.\ref{fig:Cr54ren}.

For the $N=32$ nuclei,
the shell-model energy levels of $^{58}$Fe and $^{56}$Cr
are presented in Figs.\ref{fig:Fe58eng} and \ref{fig:Cr56eng},
in comparison with the observed ones.
This shell-model calculation is performed
by using the computer code VECSSE\cite{VECSSE}.
The $k\leq 2$ space leads to the $M$-scheme dimension of
$631,670$ for $^{58}$Fe and $621,478$ for $^{56}$Cr.
Because available experimental data for these nuclei
are not as abundant as for the $N=30$ isotones,
a stringent assessment of the calculated results
appears to be difficult.
However, we see that
this shell-model calculation reproduces the observed levels 
reasonably well
also for these $N=32$ nuclei.

The structure of neutron $S$-- and $D$--pairs in $^{58}$Fe ($^{56}$Cr)
is assumed to be the same as in $^{56}$Fe ($^{54}$Cr),
for the sake of simplicity.
The $|D_\nu^2 ; J\rangle$ bases are produced 
so as to have the generalized seniority\cite{ref:gss} of four,
by removing lower-seniority components.
In $^{56}$Cr, the lack of the $|D_\pi^2;0^+ \rangle$ component
causes a modification of the $SD$ bases, as in $^{54}$Cr.
The energy levels in $^{58}$Fe and $^{56}$Cr within the $SD$ space
are shown in Figs.\ref{fig:Fe58ren} and \ref{fig:Cr56ren},
in comparison with the shell-model ones.

Figures~\ref{fig:Cr54ren}, \ref{fig:Fe58ren} and \ref{fig:Cr56ren}
indicate, respectively, that the H$^2$CM works well
for the $0^+_1$ and $2^+_1$ states of $^{54}$Cr, $^{58}$Fe and 
$^{56}$Cr, to the same extent as in $^{56}$Fe.
As shown in Table~\ref{tab:ovl},
the wavefunctions, as well as the energy levels,
are in good agreement with the realistic shell-model ones.
In $^{54}$Cr and $^{56}$Cr,
the H$^2$CM wavefunctions have more than $95\%$ overlap
with the shell-model eigenstates for $0_1^+$,
and more than $90\%$ for $2_1^+$, as in $^{56}$Fe.
These numbers are somewhat smaller in $^{58}$Fe,
originating in the smaller overlaps of the unrenormalized $SD$--states
and the shell-model eigenstates.
Though we have presumed the same neutron-pair structure 
as in $^{56}$Fe,
a different choice may improve the overlaps.
The overlaps of the H$^2$CM wavefunctions
with the shell-model ones are still large,
exceeding $90\%$ for $0_1^+$ and $80\%$ for $2_1^+$.
Thus the quadrupole collective motion in these nuclei
can plausibly be described via the H$^2$CM,
based on the realistic shell model.

Although the calculation without the renormalization
yields quite high ground-state energy,
it produces the energy spectrum similar to that in the H$^2$CM result
for most lower-lying states.
This consequence is not a trivial one,
and seems to give a rather deep insight upon the correspondence
between the $SD$--pair states and the IBM--2 states.

It should be mentioned here that
the renormalization proposed by van Egmond and Allaart\cite{ref:EA84}
has similar aspects to the present one.
They have included various correlations induced
through a nearly realistic shell-model hamiltonian.
To obtain a renormalized wavefunction of the state with one $D$--pair,
the hamiltonian is diagonalized on the one- and two-broken-pair bases.
Most of these bases are involved in the H$^1$CM,
if their work is compared with the H$^n$CM.
However, a few of them emerge only in H$^2$CM.
In this sense, a part of the H$^2$CM effect has been 
taken into account in Ref.\cite{ref:EA84}.
The H$^n$CM exploited in the present work, on the other hand,
provides us with a systematic way to pick up important bases.
Wavefunctions of higher $SD$--states are also handled by the H$^n$CM.
A significant difference from Ref.\cite{ref:EA84},
as well as from other studies,
is that the $^{56}$Ni--core excitation plays a certain role
in the present case,
which may be characteristic to the present mass region.
We should note that, because of this point,
the $\left|0^+(S^{N^{\rm B}});F=F_{\rm max}\right\rangle$ basis
is modified to an appreciable extent.

It has been demonstrated how efficient the H$^n$CM is.
Despite the relatively heavy leakage out of the $SD$ space 
($W^{(0)}$),
the H$^2$CM yields reasonable energies and wavefunctions
for the lowest-lying states.
The leakage mainly arises from the $k>0$ configurations,
namely from the $^{56}$Ni--core breaking.
It is commented that,
if we had an appropriate effective interaction in the $k=0$ space,
the H$^1$CM, or even the bare $SD$ wavefunctions,
might have worked more efficiently.

\section{IBM--2 parameters}
\label{sec:param}

We next calculate IBM--2 parameters by extending the OAI mapping,
based on the H$^n$CM.
There have been numerous investigations devoted to the derivation 
of the IBM--2 hamiltonian
from schematic interactions like a pairing-plus-quadrupole 
interaction
or a surface-delta interaction\cite{ref:DPBD,ref:Miz92,ref:Sch83}.
A semi-realistic interaction with the single-range Yukawa form
has been applied to investigate several of the IBM--2 parameters
in Ref.\cite{ref:EA84}.
On the other hand,
this is the first work to derive IBM--2 hamiltonian 
from a realistic shell-model interaction;
the Kuo--Brown interaction, in the present case.

We employ the following form of the IBM--2 hamiltonian,
including all the possible one- and two-body terms,
\begin{eqnarray}
  H^{\rm B} &=& E_0 + \sum_{\rho=\pi,\nu} \epsilon_{d_\rho} 
  \hat N_{d_\rho}
     + \sum_{\rho=\pi,\nu} V_\rho^{\rm B}
    - \kappa \hat Q_\pi \cdot \hat Q_\nu + \sum_{J=1,2,3} \xi_J 
    \hat M_J
  \nonumber \\
  &&~~+ \sum_{J=0,2,4} b_J [d_\pi^\dagger d_\nu^\dagger]^{(J)}
   \cdot [\tilde d_\pi \tilde d_\nu]^{(J)} , \label{eq:Hboson}
\end{eqnarray}
where
\begin{eqnarray}
  \hat Q_\rho &=& [d_\rho^\dagger s_\rho + s_\rho^\dagger 
  \tilde d_\rho]^{(2)}
    + \chi_\rho [d_\rho^\dagger \tilde d_\rho]^{(2)} ,\\
  V_\rho^{\rm B} &=& {1\over 2} v_{0,\rho}
   \{s_\rho^\dagger s_\rho^\dagger \cdot [\tilde d_\rho 
   \tilde d_\rho]^{(0)} + \mbox{h.c.} \}
  + {1\over{\sqrt{2}}} v_{2,\rho}
   \{d_\rho^\dagger s_\rho^\dagger \cdot [\tilde d_\rho 
   \tilde d_\rho]^{(2)} + \mbox{h.c.} \} \nonumber \\
  && + {1\over 2} \sum_{J=0,2,4} c_{J,\rho}
   [d_\rho^\dagger d_\rho^\dagger]^{(J)}
    \cdot [\tilde d_\rho \tilde d_\rho]^{(J)} .
  \label{eq:Vrho}
\end{eqnarray}
The so-called Majorana terms,
which control the energy of the mixed-symmetry states
relative to the symmetric states,
are defined by
\begin{mathletters} \label{eq:Majorana}
\begin{equation} \hat M_J = [d_\pi^\dagger d_\nu^\dagger]^{(J)} \cdot
 [\tilde d_\nu \tilde d_\pi]^{(J)}~~\mbox{(for $J=1,3$)} , 
 \end{equation}
\begin{equation} \hat M_2 = {1\over 2}
 [d_\pi^\dagger s_\nu^\dagger - s_\pi^\dagger d_\nu^\dagger]^{(2)}
    \cdot [\tilde d_\pi s_\nu - s_\pi \tilde d_\nu]^{(2)} . 
    \end{equation}
\end{mathletters}

In the OAI mapping, a boson image of a certain nucleon operator
is obtained from matrix elements concerning low-seniority states.
The parameters are fixed so that a boson matrix element 
should be equal
to the corresponding matrix element in the collective fermion space.
We here consider $F$--spin also, in addition to the seniority.
The boson matrix elements are equated to the fermion ones,
similarly to the sequence of the bases in the H$^n$CM
discussed in Section~\ref{sec:illust}.
This procedure is briefly illustrated below.

We first consider the $s$--boson condensate $|s^{N^{\rm B}})$.
Since $E_0=(s^{N^{\rm B}}|H^{\rm B}|s^{N^{\rm B}})$,
the parameter $E_0$ in Eq.(\ref{eq:Hboson}) is fixed by the equation
\begin{equation} E_0 = \left\langle 0^+(S^{N^{\rm B}}); F=F_{\rm max}
\right|H\left| 0^+(S^{N^{\rm B}}); F=F_{\rm max} \right\rangle,
\label{eq:E0} \end{equation}
where $H$ stands for the shell-model hamiltonian.
When we regard the $|0^+(S^{N^{\rm B}})\rangle$ basis
as the unrenormalized one,
we obtain the unrenormalized value of $E_0$ from Eq.(\ref{eq:E0}).
By putting the renormalized wavefunction 
for $|0^+(S^{N^{\rm B}})\rangle$,
the renormalized value of $E_0$ is evaluated.
In this procedure, the wavefunction renormalization
for the collective fermion states gives rise to a
renormalization of the IBM--2 parameter.
The $\kappa$ parameter is fixed from
the $\langle 0^+(S^{N^{\rm B}})|H|0^+(S^{N^{\rm B}-2}D^2)\rangle$
matrix element.
Then $\epsilon_{d_\pi}$ and $\epsilon_{d_\nu}$ are determined
from the following coupled equation,
\begin{mathletters}
\begin{eqnarray}
&&E_0+{1\over{N^{\rm B}}}(N_\pi^{\rm B} \epsilon_{d_\pi}
 + N_\nu^{\rm B} \epsilon_{d_\nu})
-{{2N_\pi^{\rm B} N_\nu^{\rm B}}\over{N^{\rm B}}}\kappa \nonumber \\
&&~~= \left\langle 2^+(S^{N^{\rm B}-1}D); F=F_{\rm max} \right|
H\left| 2^+(S^{N^{\rm B}-1}D); F=F_{\rm max} \right\rangle,
\end{eqnarray}
\begin{eqnarray}
&&{\sqrt{N_\pi^{\rm B} N_\nu^{\rm B}}\over{N^{\rm B}}}
(N_\pi^{\rm B} \epsilon_{d_\pi} - N_\nu^{\rm B} \epsilon_{d_\nu})
+{{\sqrt{N_\pi^{\rm B} N_\nu^{\rm B}} (N_\pi^{\rm B}-N_\nu^{\rm B})}
\over{N^{\rm B}}}\kappa \nonumber \\
&&~~= \left\langle 2^+(S^{N^{\rm B}-1}D); F=F_{\rm max} \right|
H\left| 2^+(S^{N^{\rm B}-1}D); F=F_{\rm max}-1 \right\rangle.
\end{eqnarray}
\end{mathletters}
The other parameters are evaluated in an analogous manner.
It should be noticed that in the unrenormalized case
this procedure yields the same results as the OAI mapping.

The resultant parameters,
the unrenormalized ones and the renormalized ones via the H$^2$CM,
are displayed in Table~\ref{tab:en_par}.
For neither $^{56}$Fe nor $^{54}$Cr, 
the interaction among neutron bosons
($V_\nu^{\rm B}$) has any effect, since $N_\nu^{\rm B}=1$.
Hence the parameters in $V_\nu^{\rm B}$ have not been determined
for these nuclei.
Likewise, $V_\pi^{\rm B}$ in the Fe nuclei is not given.
For the Cr nuclei,
the lack of the $|D_\pi^2; 0^+\rangle$ component is realized
by setting $c_{0,\pi}=\infty$,
while $v_{0,\pi}$ is indeterminate.

It is often assumed, in phenomenological studies,
that the IBM--2 hamiltonian is comprised
only of the second, fourth and fifth terms of Eq.(\ref{eq:Hboson}).
Note that, as far as excitation energies are concerned,
the constant term $E_0$ plays no role.
The $2_1^+$ excitation energy is governed
mainly by the $\hat N_d$ and $\hat Q_\pi \cdot \hat Q_\nu$ terms.
For the parameters associated with these terms,
the difference between the results with and without 
the renormalization is not large.
For instance, $\chi_\pi$ of the Cr nuclei vanishes 
before the renormalization,
and it remains small after the H$^2$CM.
In the $N=32$ nuclei, we have very small $\chi_\nu$
both in the unrenormalized and H$^2$CM cases.
A certain nucleus-dependence has been expected
for the $\chi$ parameters\cite{ref:OAI}.
This variation with the increasing valence nucleon number
is rapid in this region,
because the size of the shell is small, compared with heavier nuclei.

It is found that, 
while the Majorana interaction is negligibly small 
before the renormalization,
it becomes sizably repulsive due to the renormalization.
In reality, the totally symmetric states are pushed down
more than the mixed-symmetry states,
absorbing the more effect of non-$SD$ degrees of freedom.
Thereby the mixed-symmetry states are pushed up to some extent,
relative to the lowest-lying states.
This point is already viewed in the fermion spectra shown
in the preceding section.
The repulsive $\hat M_1$ term after the H$^2$CM is compatible
with increase of the excitation energy of the $1^+$ states
in Figs.\ref{fig:Fe56ren}, \ref{fig:Cr54ren}, \ref{fig:Fe58ren}
and \ref{fig:Cr56ren}.

It has been shown\cite{ref:DPBD,ref:Sch83} that,
if we calculate a boson image of a schematic interaction,
the Majorana terms emerge as a renormalization effect.
A similar effect occurs also for a realistic interaction derived 
from the $G$--matrix.
Nevertheless, there is a difference in mechanism.
Whereas only the influence of $g$--boson degree of freedom
is considered in Refs.\cite{ref:DPBD,ref:Sch83},
various correlation is included in the present renormalization.
In particular, the core excitation effect looks to play a certain role
in the present case.
Energies of the lowest $0^+$ and $2^+$ ({\it i.e.}, symmetric) states
are lowered greatly, mainly due to the coupling 
to the core excitation.
This mechanism works less for the mixed-symmetry components,
resulting in the repulsive Majorana terms.

Comparing the H$^2$CM results on the Majorana terms 
among the four nuclei,
we find certain nucleus-dependence of the parameters.
In $^{56}$Fe, $\xi_1$ is fairly large with $\xi_2$ and $\xi_3$
remaining quite small.
The other nuclei have positive values for all of these parameters.
The $\xi_2$ parameter is somewhat smaller than the others.

It is also noticed that $V_\rho^{\rm B}$
and the last term of Eq.(\ref{eq:Hboson})
are not negligibly small.
Though they hardly contribute to the lowest-lying states 
in these nuclei,
some higher-lying states are affected to a certain extent.

By diagonalizing the IBM--2 hamiltonian,
we obtain the energy levels within the IBM--2.
They are already displayed in Figs.\ref{fig:Fe56ren}, 
\ref{fig:Cr54ren}, \ref{fig:Fe58ren} and \ref{fig:Cr56ren},
by using the parameters after the H$^2$CM.
Since we include all the one- and two-body terms 
in the boson hamiltonian,
the IBM--2 levels in $^{56}$Fe, where $N^{\rm B}=2$,
are exactly the same as those of the collective fermion space.
In $^{54}$Cr, the boson energy levels are very close
to those of the collective fermion space,
while the perfect agreement can be made
if we use three-body terms in the boson hamiltonian.
This indicates that the boson many-body terms are not important.
The same holds for $^{58}$Fe and $^{56}$Cr.

We next turn to electromagnetic transition operators.
The following shell-model $E2$ operator is assumed,
\begin{equation} T(E2) = \sum_{\rho=\pi,\nu}
 e_\rho^{\rm eff} \sum_{i\in\rho} r_i^2 Y^{(2)}(\hat{\bf r}_i),
\end{equation}
with $e_\pi^{\rm eff}=1.4e$ and $e_\nu^{\rm eff}=0.9e$.
The single-particle matrix-elements are evaluated
by the harmonic-oscillator wavefunctions with $b=56^{1/6}=1.956$fm.
For the $M1$ operator,
\begin{equation} T(M1) = \sqrt{{3\over{4\pi}}} \sum_{\rho=\pi,\nu}
\left\{ g_{l,\rho}^{\rm eff} \sum_{i\in\rho} l_i
 + g_{s,\rho}^{\rm eff} \sum_{i\in\rho} s_i \right\}, \end{equation}
and we take $g_{l,\rho}^{\rm eff}=g_{l,\rho}^{\rm free}$,
$g_{s,\rho}^{\rm eff}=0.5g_{s,\rho}^{\rm free}$.
These electromagnetic operators are the same as 
in Ref.\cite{ref:NOS91}.
While medium effect on the $M3$ operator has not been explored 
sufficiently,
there is an evidence in the {\em sd}--shell that 
no quenching is necessary
to describe $M3$ transitions\cite{ref:bare-M3}.
The shell-model $M3$ operator is taken to be equal
to the bare-nucleon operator,
\begin{equation} T(M3) = {\sqrt{21}\over{2}} \sum_{\rho=\pi,\nu}
\left\{ g_{l,\rho}^{\rm free} \sum_{i\in\rho}
 r_i^2 [Y^{(2)}(\hat{\bf r}_i) l_i]^{(3)}
 + 2g_{s,\rho}^{\rm free} \sum_{i\in\rho}
  r_i^2 [Y^{(2)}(\hat{\bf r}_i) s_i]^{(3)} \right\}, \end{equation}
where the $g$-parameter are the same as for the $M1$ operator.

The IBM--2 operators are
\begin{eqnarray}
  T^{\rm B} (E2) &=& \sum_{\rho=\pi,\nu} e_\rho^{\rm B} \left\{
    [d_\rho^\dagger s_\rho + s_\rho^\dagger \tilde d_\rho]^{(2)}
    + \chi'_\rho [d_\rho^\dagger \tilde d_\rho]^{(2)} \right\} ,\\
  T^{\rm B} (M1) &=& \sqrt{{3\over{4\pi}}} \sum_{\rho=\pi,\nu}
   g_\rho^{\rm B} \hat J_\rho^{\rm B};~~
  \hat J_\rho^{\rm B} = \sqrt{10} [d_\rho^\dagger \tilde d_\rho]^{(1)}
   ,\\
  T^{\rm B} (M3) &=& \sum_{\rho=\pi,\nu} \beta_{3,\rho}
   [d_\rho^\dagger \tilde d_\rho]^{(3)} .
\end{eqnarray}
The parameters introduced above are evaluated from the matrix elements
within the collective fermion space,
analogously to the mapping for the hamiltonian.
The resultant IBM--2 parameters are shown in Table~\ref{tab:tr_par}.

The renormalization enhances the boson effective-charges 
($e_\rho^{\rm B}$),
gaining more quadrupole collectivity.
In phenomenological studies,
$e_\pi^{\rm B}=e_\nu^{\rm B}$ is sometimes postulated.
This is supported by the present microscopic study.
Though the shell-model effective-charge is smaller for neutrons
than for protons,
the neutrons have more quadrupole collectivity
because the size of the valence shell is bigger.
As a consequence, $e_\pi^{\rm B}$ and $e_\nu^{\rm B}$ are not 
so different.
There is no apparent reason for the $\chi'$ parameters to agree
with the $\chi$ parameters which appear in the boson hamiltonian,
because we adopt a realistic shell-model interaction,
not a schematic proton-neutron interaction like $Q_\pi \cdot Q_\nu$.
Nevertheless, $\chi'$ is close to $\chi$ in any case.
It has been expected that $g_\pi^{\rm B}\simeq g_{l,\pi}^{\rm free}=1$
and $g_\nu^{\rm B}\simeq g_{l,\nu}^{\rm free}=0$,
since the nucleon-spin degrees of freedom are not so active
in the quadrupole collective states.
However, $g_\pi^{\rm B}$ is certainly larger than unity
in the result without the renormalization.
This happens because the proton $SD$--pairs consist
of the single-$j$ orbit of $0f_{7/2}$.
On the contrary, when we carry out the renormalization,
the excitation from $0f_{7/2}$ to $0f_{5/2}$ leads
towards saturation of the nucleon spin.
Therefore the $g_\pi^{\rm B}$ value is reduced
until $g_\pi^{\rm B}\simeq 1$ is restored to a good extent.
The quenching of $\beta_{3,\pi}$ due to the renormalization
is explained by a similar spin-saturation mechanism.
It is somewhat surprising that the influence of the renormalization
on $g_\nu^{\rm B}$ and $\beta_{3,\nu}$ is so small.
This will be partly because the spin saturation occurs
already in the unrenormalized neutron parameters,
resulting in almost vanishing values.

The same quenching mechanism for the magnetic transition parameters
will prevail in heavier nuclei,
where a unique-parity orbit is involved in the valence shell.
Since the spin-orbit partner of the unique-parity orbit
is absent in the valence shell,
the nucleon-spin content could influence
$g_\rho^{\rm B}$ and $\beta_{3,\rho}$ to a considerable extent,
when we ignore effects of excitation across major shells.
However, by taking into account the excitation effects,
those parameters will be quenched,
owing to the spin-saturation tendency\cite{ref:ASBH} in the dynamics.
It is expected for the $M1$ parameters
that $g_\pi^{\rm B}$ approaches unity and $g_\nu^{\rm B}$ 
almost vanishes.

The dependence of the IBM--2 parameters on valence nucleon numbers
is accounted for, in most cases, in terms of the quasi-spin properties
of the relevant nucleon operator\cite{ref:OAI}.
The $E2$ operator behaves as a vector 
in the quasi-spin space\cite{ref:AI66}.
Weak dependence is suggested for $e_\rho^{\rm B}$,
and is confirmed in Table~\ref{tab:tr_par}.
The number-dependence of $\chi'$ is strong, as is expected.
So is $\chi$ in the boson hamiltonian.
Since $T(M1)$ and $T(M3)$ are quasi-spin scalars\cite{ref:AI66},
the $g_\rho^{\rm B}$ and $\beta_{3,\rho}$ parameters
are expected to be nearly constant.
This is true for $g_\pi^{\rm B}$, $g_\nu^{\rm B}$ and $\beta_{3,\pi}$,
but a considerable deviation is seen in $\beta_{3,\nu}$.
This is a sort of many-body effect,
originating in the subshell structure.

\section{Discussion on H\lowercase{$^n$}CM}
\label{sec:discuss}

\subsection{Choice of primary bases}
\label{subsec:basis}

In this section, we return to discussion on the H$^n$CM.

Even when the subspace $W_J^{(0)}$ is fixed,
there still remains an ambiguity in choosing 
the primary orthonormal bases
$\left\{ \Psi_1^{(0)}, \Psi_2^{(0)}, \cdots, \Psi_l^{(0)} \right\}$.
The bases can be changed by a unitary transformation.
Since the couplings between $\phi_\lambda^{(\nu-1)}$
and $\phi_\lambda^{(\nu)}$
are not uniform for various $\lambda$,
the unitary transformation may lead 
to a different renormalized basis-set in the H$^n$CM.
Moreover, because of the orthogonalization stated 
in Section~\ref{sec:illust},
the H$^n$CM bases generally depends on the ordering 
of the primary bases.

The following three choices will be possible:
\begin{enumerate}
\item Some orthonormal basis-set is postulated
for $\Psi_\lambda^{(0)}$ by a physical insight. \label{itm:bas1}
\item The eigenstates within $W_J^{(0)}$ are taken 
as $\Psi_\lambda^{(0)}$.
They are placed in order according to the eigenenergies.
\label{itm:bas2}
\item $\Psi_\lambda^{(0)}$ is redefined for each step $n$,
so that $\Psi_\lambda^{(n-1)}$ should be an eigenstate 
within $W_J^{(n-1)}$.
They are put in order according to the eigenenergies. \label{itm:bas3}
\end{enumerate}
To fulfill the space closure of Eq.(\ref{eq:Vclose}),
the couplings between renormalized states
and remaining degrees of freedom
should be small.
From this viewpoint, the latter choice seems favorable.
On the other hand, the latter requires more complication
in the numerical treatment.

The basic dynamical properties should be well represented
by the primary bases,
otherwise the renormalization does not converge with small $n$.
In the practical case of the Cr--Fe nuclei
in Sections~\ref{sec:illust} and \ref{sec:result},
we have adopted (\ref{itm:bas1}),
with the $U_{\pi+\nu}(5)\otimes SU_F(2)$ bases 
of IBM--2\cite{ref:IBM}.
This will be appropriate because those nuclei seem 
to be nearly spherical.
In deformed region, another choice might be better\cite{ref:Yos94}.

\subsection{Choice of renormalized bases}
\label{subsec:frag}

One could pursue the convergence of H$^n$CM,
by increasing the power $n$.
However, besides tedious computations,
it might lead to a dissimilar wavefunction of 
$\Psi_\lambda^{(n)}$
from that of $\Psi_\lambda^{(0)}$.
Then it is not reasonable to regard $\Psi_\lambda^{(n)}$
as a renormalized state.
In some cases, this is circumvented
if an eigenstate in $\Gamma_\lambda^{(n)}$
having the largest overlap with $\Psi_\lambda^{(0)}$
is adopted as $\Psi_\lambda^{(n)}$,
instead of the lowest-lying one.

A complication occurs when there is a substantial fragmentation
of the state which carries the main character of the primary state.
As has been pointed out in Ref.\cite{ref:NOS91},
the mixed-symmetry $2^+$ state of $^{56}$Fe may be the case.
The $2^+_2$ and $2^+_4$ states share appreciable fractions
of this collective component.
As $n$ increases,
the diagonalization within $\Gamma_\lambda^{(n)}$
will lead to a problem at some value of $n$;
plural eigenstates have considerable amplitudes
of the primary basis $\Psi_\lambda^{(0)}$.
It will not be desirable, in such a case,
to choose a single eigenstate in $\Gamma_\lambda^{(n)}$
as a renormalized basis $\Psi_\lambda^{(n)}$.

A solution to this problem is to adopt a linear combination
of the few eigenstates in $\Gamma_\lambda^{(n)}$
as a renormalized basis.
We can set a criterion of minimum amplitude for the states 
to be included.
Another practical choice is just stopping at a certain $n$.
It should be emphasized that, in any case,
monitoring the H$^n$CM outcome for each step is significant.

In the actual case of the Cr--Fe nuclei,
we do not come across the problems stated above,
up to the H$^2$CM.
The lowest-lying eigenstate in $\Gamma_\lambda^{(n)}$
has the largest overlap with $\Psi_\lambda^{(0)}$,
and no serious fragmentation is viewed.
As shown in Section~\ref{sec:result},
the convergence in $0^+_1$ and $2^+_1$ is so rapid
that we could acquire a good approximation by the H$^2$CM.
In this respect, the H$^2$CM seems good enough
to investigate collective states of the Cr--Fe nuclei.

\subsection{H\lowercase{$^n$}CM and Lanczos method}
\label{subsec:Lanczos}

As has been mentioned earlier,
there is a common part between the H$^n$CM
and the Lanczos diagonalization method.

The H$^n$CM energy levels are obtained 
via two steps of diagonalization;
one within the subspace $\Gamma_\lambda^{(n)}$
and the other within $W_J^{(n)}$.
The dimension of $\Gamma_\lambda^{(n)}$ is $(n+1)$,
while that of $W_J^{(n)}$ is $l$.
The basis production and the diagonalization 
within $\Gamma_\lambda^{(n)}$
is similar to the Lanczos method.
In the case that there is a single basis in $W_J^{(0)}$,
the H$^n$CM procedure is the same as the Lanczos method
starting from $\Psi^{(0)}$,
since we do not need the diagonalization within $W_J^{(n)}$.
In other cases, the orthogonalization
between $\Gamma_\lambda^{(n)}$ and $\Gamma_{\lambda'}^{(n)}$
$(\lambda\ne\lambda')$ in the H$^n$CM
does not appear in the Lanczos method.
Apart from this difference, the H$^n$CM is exploited
so as to make good use of the advantage of the Lanczos method.

An emphasis should be put on the primary bases:
we have requested that they should have a simple structure
but carry the basic dynamics of the system.
For instance, the $SD$ states are taken as the primary bases
in the application to the Cr--Fe nuclei
in Section~\ref{sec:result}.
Because of these properties, it is expected in many cases that
the H$^n$CM is more efficient than the Lanczos method.
The number of bases in the H$^n$CM is given by $l(n+1)$
for each $J$, which can be smaller than
that necessary in the Lanczos method.
In practice, within a fixed $J$,
even less than 10 bases yield good accuracy
for the lowest-lying levels of the Cr--Fe nuclei via the H$^n$CM,
whereas about 50 bases are normally required in the Lanczos method.
It is noted that each diagonalization is performed
for a matrix with quite a small dimension, $(n+1)$ or $l$.

\subsection{H\lowercase{$^n$}CM and a perturbative renormalization}
\label{subsec:perturb}

In this subsection, the H$^n$CM is discussed
in connection with a perturbative method of renormalization.
A more detailed discussion is given in Ref.\cite{ref:Nak91}.

Let us recall Feshbach's projection method\cite{ref:Fes62},
a well-known method of incorporating truncation effects.
We define the $P$--space as the space of the primary bases;
the original $SD$ space in the present case.
The total space corresponds to the $k\leq 2$ shell-model space,
while the original hamiltonian is the Kuo--Brown hamiltonian.
We shall take into account the effects of the $Q$--space,
which is spanned by the non-primary bases.
Note that all the $\phi$-bases of Eq.(\ref{eq:phi})
belong to the $Q$--space.
The projection operator onto the $P$--space is denoted by $\hat P$,
and that onto the $Q$--space by $\hat Q$, 
namely $\hat Q = 1 - \hat P$.

In Feshbach's method,
the renormalized $P$--space hamiltonian is given by
\begin{equation} {\tilde H}_P = H_P
 + \hat P H \hat Q {1\over{E-H_Q}}\hat Q H\hat P,
\label{eqC:renormH} \end{equation}
where
\begin{equation} H_P=\hat P H\hat P,~~H_Q=\hat Q H\hat Q . 
\end{equation}
For the exact treatment of $H_Q$,
eigenvalues of $H_Q$ have to be calculated.
Let an eigenstate of $H_Q$ be denoted by $|q_i\rangle$.
Notice that $\hat Q|q_i\rangle = |q_i\rangle$
and $\hat Q = \sum_i |q_i\rangle\ \langle q_i|$.
Then the second term on the right-hand side of Eq.(\ref{eqC:renormH})
is rewritten as
\begin{equation} \sum_i \hat P H|q_i\rangle {1\over{E-E(q_i)}}
 \langle q_i|H\hat P, \end{equation}
where $E(q_i) = \langle q_i |H|q_i\rangle$.
Though Eq.(\ref{eqC:renormH}) gives an exact way
to incorporate the influence of the $Q$--space into the hamiltonian,
it is difficult and not advantageous to handle
without any approximation in most cases,
because of the following two reasons.
If one wish to know exact eigenenergies,
$H_Q$ must be treated exactly, which is
usually a matrix with enormous dimension.
Moreover, a non-linear coupled equation must be solved,
since the eigenenergy $E$ is also contained
in the denominator of the second term.

For the sake of simplicity,
our discussion is restricted to a fixed $J$ 
(conserved quantum number),
without loss of generality.
We introduce the following state generated from $\Psi_\lambda^{(0)}$,
\begin{equation} |\overline{\phi}_\lambda \rangle
 = \sum_i x_{i,\lambda} |q_i\rangle
\propto \hat Q H\left|\Psi_\lambda^{(0)}\right\rangle,~~
(\lambda = 1,2,\cdots,l) \label{eq:qa} \end{equation}
where
\begin{equation} x_{i,\lambda} = {{\langle q_i|H|\Psi_\lambda^{(0)}
\rangle}
\over{\sqrt{\sum_{i'} \left|\langle q_{i'}|
H|\Psi_\lambda^{(0)}\rangle\right|^2}}}
 = {{\langle q_i|H|\Psi_\lambda^{(0)}\rangle}\over
{\sqrt{\left\langle\Psi_\lambda^{(0)}\right|H \hat Q H
\left|\Psi_\lambda^{(0)}\right\rangle}}}. \end{equation}
It is noticed that, with the notation in Section~\ref{sec:illust},
$\overline{\phi}_\lambda$ can be expressed
as $P^{\cal O}_{\{\Psi_1^{(0)},\Psi_2^{(0)},\cdots,\Psi_l^{(0)}\}}
\cdot H\Psi_\lambda^{(0)}$,
which is different from the basis $\phi_\lambda^{(1)}$
only in the lack of orthogonalization 
between $\overline{\phi}_\lambda$
and $\overline{\phi}_{\lambda'}$ $(\lambda\ne\lambda')$.

We now substitute a $c$-number
$E(\overline{\phi}_\lambda)
=\langle \overline{\phi}_\lambda |H|\overline{\phi}_\lambda\rangle$
for $H_Q$ in Eq.(\ref{eqC:renormH}).
Then a diagonal matrix element of $\tilde H_P$ is approximated by
\begin{eqnarray} \left\langle\Psi_\lambda^{(0)}\right|\tilde H_P
 \left|\Psi_\lambda^{(0)}\right\rangle
&\simeq& \langle \Psi_\lambda^{(0)}|H|\Psi_\lambda^{(0)}\rangle
 + \langle\Psi_\lambda^{(0)}|H|\overline{\phi}_\lambda\rangle
 {1\over{E-E(\overline{\phi}_\lambda)}}
 \langle \overline{\phi}_\lambda |H|\Psi_\lambda^{(0)}\rangle
  \nonumber \\
&=& \langle \Psi_\lambda^{(0)}|H|\Psi_\lambda^{(0)}\rangle
 + {1\over{E-E(\overline{\phi}_\lambda)}} \langle\Psi_\lambda^{(0)}|
 H \hat Q H|\Psi_\lambda^{(0)}\rangle .
\label{eq5:H'c} \end{eqnarray}
This is a kind of closure approximation,
since it is given by replacing the energy denominator by a $c$-number.
It is not easy, in general, to evaluate $E$
in the energy denominator properly.
By substituting unperturbed energy $E_\lambda^{(0)}
=\left\langle\Psi_\lambda^{(0)}\right| H
\left|\Psi_\lambda^{(0)}\right\rangle$
for it,
Eq.(\ref{eq5:H'c}) becomes equivalent 
to the second-order perturbation
combined with the closure approximation.
It is remarked that, however, the unperturbed energy is too high
in the practical case of the Cr--Fe nuclei,
causing too big renormalization effect.
If we neglect the non-orthogonality between $\overline{\phi}_\lambda$
and $\overline{\phi}_{\lambda'}$ for $\lambda\ne\lambda'$,
the space $\Gamma_\lambda^{(1)}$ becomes
$\left\{\Psi_\lambda^{(0)}, \overline{\phi}_\lambda \right\}$.
In addition, if $E$ is estimated by diagonalizing the hamiltonian
in this $\Gamma_\lambda^{(1)}$ space,
we obtain $E_\lambda^{(1)}$.
Then the right-hand side of Eq.(\ref{eq5:H'c}) is equivalent
to the diagonal element of the H$^1$CM collective hamiltonian.
Note that $E_\lambda^{(1)}$ is lower than $E_\lambda^{(0)}$,
and is closer to the exact $E$ for low-lying states.

Observing the above relation between the H$^1$CM
and the perturbative renormalization with closure approximation,
we can claim that the H$^1$CM is
an improvement from the perturbative method
in the following points;
(i) the over-counting arising from the non-orthogonality
between $\overline{\phi}_\lambda$ and $\overline{\phi}_{\lambda'}$
$(\lambda\ne\lambda')$ is removed,
(ii) off-diagonal elements are evaluated
in a consistent manner with diagonal ones,
and (iii) $E$ in the energy denominator is improved.
If the overlap between $\overline{\phi}_\lambda$
and $\overline{\phi}_{\lambda'}$ is negligible,
which somewhat depends on how to choose the original basis-set
$\left\{\Psi_\lambda^{(0)}; \lambda=1,2,\cdots,l \right\}$,
$\overline{\phi}_\lambda$ and $\phi_\lambda^{(1)}$
becomes quite similar.
In order to further make the relationship between the two methods 
more transparent,
we shall restrict ourselves to the diagonal elements
and ignore how $E$ is estimated.

If the second term of the right-hand side of Eq.(\ref{eqC:renormH})
is expanded by the parameter
\begin{equation} \zeta_{i,\lambda}
 = {{E(\overline{\phi}_\lambda)-E(q_i)}\over{E-
 E(\overline{\phi}_\lambda)}}, \end{equation}
we obtain\cite{ref:Nak91}
\begin{eqnarray}
&&\left\langle\Psi_\lambda^{(0)} \right|\hat P H\hat Q
 {1\over{E-H_Q}} \hat Q H\hat P\left|\Psi_\lambda^{(0)} \right\rangle
  \nonumber\\
&&~~~= {1\over{E-E(\overline{\phi}_\lambda)}}
 \left\langle\Psi_\lambda^{(0)} \right| H \hat Q H
 \left|\Psi_\lambda^{(0)} \right\rangle
 \left\{1+\left[{{\sigma_\lambda (H_Q)}
 \over{E-E(\overline{\phi}_\lambda)}}\right]^2 \right\}
+ O(\zeta^3) , \label{eqC:cancel0}
\end{eqnarray}
where
\begin{equation} [\sigma_\lambda (H_Q)]^2
 = \langle \overline{\phi}_\lambda |H_Q^2|\overline{\phi}_\lambda
  \rangle
 - [E(\overline{\phi}_\lambda)]^2 \label{eqC:sigma} \end{equation}
represents variance of $H_Q$ in the state $\overline{\phi}_\lambda$.
In a similar manner,
the $O(\zeta^n)$ term corresponds
to correction due to the $n$-th moment
of the distribution of $|q_i\rangle$'s.
By comparing Eq.(\ref{eqC:cancel0}) with Eq.(\ref{eq5:H'c}),
it is found in $O(\zeta^2)$ that the distribution
of $\overline{\phi}_\lambda$ over the eigenstates of $H_Q$
generally enhances the effect of the renormalization on energies.

The above discussion is useful
to acquire an intuitive picture of the H$^n$CM.
The terms regarding $H_Q^2$ are fully taken into account 
in the H$^2$CM.
Therefore, as far as diagonal elements are concerned,
the difference between Eq.(\ref{eqC:cancel0}) and the H$^2$CM 
is only in $O(\zeta^3)$.
In comparison with the H$^1$CM,
an advantage of the H$^2$CM is the inclusion
of the $O(\zeta^2)$ effect\cite{ref:Nak91}.
As stated already, in the perturbative theory,
which is connected to the H$^1$CM well,
the coupling of a primary state with the outer space ($Q$--space)
is treated by using an averaged energy of the non-primary states
$E(\overline{\phi}_\lambda)$, ignoring their distribution.
The second-order effect is, in essence, the correction
due to the distribution of the coupled states 
in terms of the variance,
as shown in Eq.(\ref{eqC:cancel0}).
By extending the present discussion to higher order,
it turns out that the $n$-th order effect of H$^n$CM 
essentially corresponds to the $n$-th moment of the distribution
of $\overline{\phi}_\lambda$ in the $Q$--space.

We now look back at the results shown
in Figs.\ref{fig:Fe56ren}, \ref{fig:Cr54ren},
\ref{fig:Fe58ren} and \ref{fig:Cr56ren} in Section~\ref{sec:result}.
In proceeding from H$^1$CM to H$^2$CM,
the higher-lying states tend to go down more sharply.
This implies that the variance $\sigma_\lambda(H_Q)$
is more important in the higher-lying state.
On the other hand, some mixed-symmetry states 
with relatively low energy
(for instance, the lowest $1^+$ and the second $2^+$ states
in the collective space of each nucleus)
do not come down so rapidly,
compared with their surrounding states.
Relatively small $\sigma_\lambda(H_Q)$ is suggested
for those mixed-symmetry degrees of freedom,
giving rise to the repulsive Majorana interaction
in the IBM--2 hamiltonian.

\section{Summary}
\label{sec:summary}

In order to study quadrupole collective modes
based on a realistic shell model,
we develop the $H^n$--cooling method (H$^n$CM),
which leads to a wavefunction renormalization
by incorporating the effect of the dynamical correlation.
In practice, the H$^n$CM is applied to the $SD$ space 
of the Cr--Fe nuclei;
$^{56}$Fe, $^{54}$Cr, $^{58}$Fe and $^{56}$Cr.
While the shell-model ground-state wavefunction is not fully covered
with the simple $SD$--pair degrees of freedom,
the shell-model $0_1^+$ and $2_1^+$ energies and wavefunctions
are nicely approximated by considering up to 
the second power of $H$ (H$^2$CM).
On the other hand, as far as the energy difference is concerned,
the excitation spectra after the H$^2$CM do not differ very much
from those without the renormalization.
Note that the $^{56}$Ni--core excitation is taken into account
in the present renormalization,
as well as some effect of other like-nucleon pairs.

An extended OAI mapping is also developed and applied 
to the Cr--Fe nuclei.
This is the first work to evaluate the IBM--2 parameters
from a realistic shell-model hamiltonian.
The wavefunction renormalization is converted
to a renormalization of the IBM--2 parameters.
Some effects of the renormalization are discussed.
Although most parameters in the IBM--2 hamiltonian
do not change considerably,
the Majorana interaction becomes sizably repulsive
as a renormalization effect.
It is indicated that many-body terms are unnecessary
in the IBM--2 hamiltonian.
In the transition operators,
the H$^n$CM gives rise to spin quenching
for the $M1$ and $M3$ proton parameters,
as well as $E2$ effective-charge enhancement.
The $\chi$ parameters in the $E2$ operator are shown
to take close values to those in the hamiltonian.
This situation is not influenced by the renormalization.

\acknowledgments
The authors are grateful to Prof. A. Gelberg
for careful reading the manuscript.
One of the authors (H. N.) thank Prof. T. Sebe
for his advice on the computer programs.

\clearpage
\begin{table}
\centering
\caption{Overlaps of wavefunctions of lowest-lying collective states,
before and after the renormalization via the H$^2$CM,
with those of the shell-model eigenstates (\%).
\label{tab:ovl}}
\begin{tabular}{c|rr|rr}
    Nucleus & \multicolumn{2}{c|}{$0_1^+$} & \multicolumn{2}{c}
    {$2_1^+$} \\
    & \multicolumn{1}{c}{unren.} & \multicolumn{1}{c|}{H$^2$CM}
 & \multicolumn{1}{c}{unren.} & \multicolumn{1}{c}{H$^2$CM} \\
   \hline
     $^{56}$Fe & $55.6$ & $97.0$ & $49.8$ & $93.3$ \\
     $^{54}$Cr & $53.5$ & $95.8$ & $47.5$ & $92.3$ \\
     $^{58}$Fe & $46.0$ & $91.2$ & $36.8$ & $83.6$ \\
     $^{56}$Cr & $54.6$ & $95.1$ & $47.4$ & $90.9$ \\
\end{tabular}
\end{table}

\begin{table}
\centering
\caption{Parameters for IBM--2 hamiltonian
derived from the Kuo--Brown shell-model hamiltonian.
\label{tab:en_par}}
\begin{tabular}{cc|rr|rr|rr|rr}
    \multicolumn{2}{c|}{Parameter} & \multicolumn{2}{c|}{$^{56}$Fe}
 & \multicolumn{2}{c|}{$^{54}$Cr} & \multicolumn{2}{c|}{$^{58}$Fe}
 & \multicolumn{2}{c}{$^{56}$Cr} \\
    && \multicolumn{1}{c}{unren.} & \multicolumn{1}{c|}{H$^2$CM}
 & \multicolumn{1}{c}{unren.} & \multicolumn{1}{c|}{H$^2$CM}
 & \multicolumn{1}{c}{unren.} & \multicolumn{1}{c|}{H$^2$CM}
 & \multicolumn{1}{c}{unren.} & \multicolumn{1}{c}{H$^2$CM} \\
   \hline
     $\epsilon_{d_\pi}$ & (MeV) & $1.022$ & $1.362$
 & $1.022$ & $1.671$ & $1.022$ & $1.216$ & $1.022$ & $1.591$ \\
     $\epsilon_{d_\nu}$ & (MeV) & $1.426$ & $1.441$
 & $1.482$ & $1.270$ & $1.888$ & $2.004$ & $2.178$ & $2.143$ \\
     $\kappa$ & (MeV) & $0.839$ & $0.926$
 & $0.679$ & $0.765$ & $0.696$ & $0.710$ & $0.558$ & $0.538$ \\
     $\chi_\pi$ && $-0.933$ & $-1.220$ & $0.000$ & $-0.202$
 & $-0.933$ & $-1.194$ & $0.000$ & $-0.164$ \\
     $\chi_\nu$ && $-1.250$ & $-1.099$ & $-1.239$ & $-1.150$
 & $-0.005$ & $0.126$ & $0.013$ & $0.205$ \\
     $\xi_1$ & (MeV) & $-0.065$ & $0.303$
 & $0.084$ & $0.186$ & $-0.043$ & $0.135$ & $0.080$ & $0.192$ \\
     $\xi_2$ & (MeV) & $0.000$ & $-0.021$
 & $0.000$ & $0.127$ & $0.000$ & $0.095$ & $0.000$ & $0.098$ \\
     $\xi_3$ & (MeV) & $-0.009$ & $0.061$
 & $0.008$ & $0.171$ & $-0.031$ & $0.198$ & $-0.020$ & $0.152$ \\
     $b_0$ & (MeV) & $0.330$ & $0.575$
 & $0.123$ & $-0.099$ & $0.565$ & $0.212$ & $0.337$ & $-0.119$ \\
     $b_2$ & (MeV) & $0.101$ & $0.320$
 & $0.034$ & $0.100$ & $0.006$ & $-0.076$ & $-0.001$ & $-0.187$ \\
     $b_4$ & (MeV) & $-0.064$ & $-0.306$
 & $-0.104$ & $-0.130$ & $0.003$ & $-0.197$ & $-0.042$ & $-0.030$ \\
     $v_{2,\pi}$ & (MeV) &---&---& $0.000$ & $0.012$ &---&---&
      $0.000$ & $0.060$ \\
     $c_{2,\pi}$ & (MeV) &---&---& $0.105$ & $-0.155$ &---&---&
      $0.100$ & $-0.020$ \\
     $c_{4,\pi}$ & (MeV) &---&---& $-0.424$ & $-1.215$ &---&---&
      $-0.469$ & $-1.033$ \\
     $v_{0,\nu}$ & (MeV) &---&---&---&---& $0.082$ & $0.063$ &
 $0.141$ & $0.233$ \\
     $v_{2,\nu}$ & (MeV) &---&---&---&---& $-0.441$ & $-0.287$ &
 $-0.615$ & $-0.690$ \\
     $c_{0,\nu}$ & (MeV) &---&---&---&---& $2.066$ & $1.609$ &
 $3.399$ & $2.439$ \\
     $c_{2,\nu}$ & (MeV) &---&---&---&---& $-0.308$ & $-1.347$ &
 $0.053$ & $-0.924$ \\
     $c_{4,\nu}$ & (MeV) &---&---&---&---& $0.447$ & $0.170$&
 $0.694$ & $0.593$ \\
\end{tabular}
\end{table}

\begin{table}
\centering
\caption{Parameters for IBM--2 transition operators
derived from the shell-model operators.
\label{tab:tr_par}}
\begin{tabular}{cc|rr|rr|rr|rr}
    \multicolumn{2}{c|}{Parameter} & \multicolumn{2}{c|}{$^{56}$Fe} 
    & \multicolumn{2}{c|}{$^{54}$Cr}
& \multicolumn{2}{c|}{$^{58}$Fe} & \multicolumn{2}{c}{$^{56}$Cr} \\
    && \multicolumn{1}{c}{unren.} & \multicolumn{1}{c|}{H$^2$CM}
    & \multicolumn{1}{c}{unren.} & \multicolumn{1}{c|}{H$^2$CM}
    & \multicolumn{1}{c}{unren.} & \multicolumn{1}{c|}{H$^2$CM}
    & \multicolumn{1}{c}{unren.} & \multicolumn{1}{c}{H$^2$CM} \\
   \hline
     $e^{\rm B}_\pi$ & ($e {\rm fm}^2$) & $6.635$ & $7.958$ & $5.418$
      & $6.563$
 & $6.635$ & $7.609$ & $5.418$ & $6.243$ \\
     $e^{\rm B}_\nu$ & ($e {\rm fm}^2$) & $6.034$ & $7.273$ & $6.009$
     & $6.618$
 & $5.057$ & $6.037$ & $5.004$ & $5.365$ \\
     $\chi'_\pi$ && $-0.933$ & $-1.127$ & $0.000$ & $-0.336$
 & $-0.993$ & $-1.178$ & $0.000$ & $-0.331$ \\
     $\chi'_\nu$ && $-1.226$ & $-1.307$ & $-1.205$ & $-1.334$
 & $0.011$ & $-0.049$ & $0.035$ & $0.010$ \\
   \hline
     $g^{\rm B}_\pi$ & ($\mu_N$) & $1.256$ & $1.110$ & $1.256$
      & $1.110$
 & $1.256$ & $1.111$ & $1.256$ & $1.110$ \\
     $g^{\rm B}_\nu$ & ($\mu_N$) & $-0.027$ & $-0.037$ & $-0.039$
      & $-0.080$
 & $-0.035$ & $-0.015$ & $-0.038$ & $-0.042$ \\
   \hline
     $\beta_{3,\pi}$ & ($\mu_N {\rm fm}^2$) & $69.1$ & $51.5$
      & $69.1$ & $58.4$
 & $69.1$ & $51.4$ & $69.1$ & $59.0$ \\
     $\beta_{3,\nu}$ & ($\mu_N {\rm fm}^2$) & $-18.8$ & $-20.7$
      & $-22.5$ & $-26.3$
 & $6.1$ & $8.6$ & $5.3$ & $10.1$ \\
\end{tabular}
\end{table}

\clearpage
\begin{figure}
\caption{\label{fig:Fe56ren}
Energy levels in the collective space
(without renormalization, with renormalization via the H$^1$CM,
and with renormalization via the H$^2$CM),
in comparison with the shell-model ones in $^{56}$Fe:
(a) Lowest $0^+$ and $2^+$ energy eigenvalues in each space,
relative to the shell-model ground-state energy.
(b) Energies relative to the lowest $0^+$ level in each space.
}
\end{figure}
\begin{figure}
\caption{\label{fig:Cr54ren}
Energy levels in the collective space
compared with the shell-model ones in $^{54}$Cr.
Energy levels obtained from the IBM--2 hamiltonian are also shown.
}
\end{figure}
\begin{figure}
\caption{\label{fig:Fe58eng}
Energy levels of $^{58}$Fe.
The experimental data are taken from Ref.\protect\cite{ref:NDS58}.
The calculated energy levels are obtained
by the $k\leq 2$ shell-model calculation
with the Kuo--Brown hamiltonian.
}
\end{figure}
\begin{figure}
\caption{\label{fig:Cr56eng}
Energy levels of $^{56}$Cr.
The experimental data are taken from Ref.\protect\cite{ref:NDS56}.
The calculated energy levels are obtained
by the $k\leq 2$ shell-model calculation
with the Kuo--Brown hamiltonian.
}
\end{figure}
\begin{figure}
\caption{\label{fig:Fe58ren}
Energy levels in the collective space,
as well as the IBM--2 energy levels,
compared with the shell-model ones in $^{58}$Fe.
}
\end{figure}
\begin{figure}
\caption{\label{fig:Cr56ren}
Energy levels in the collective space,
as well as the IBM--2 energy levels,
compared with the shell-model ones in $^{56}$Cr.
}
\end{figure}

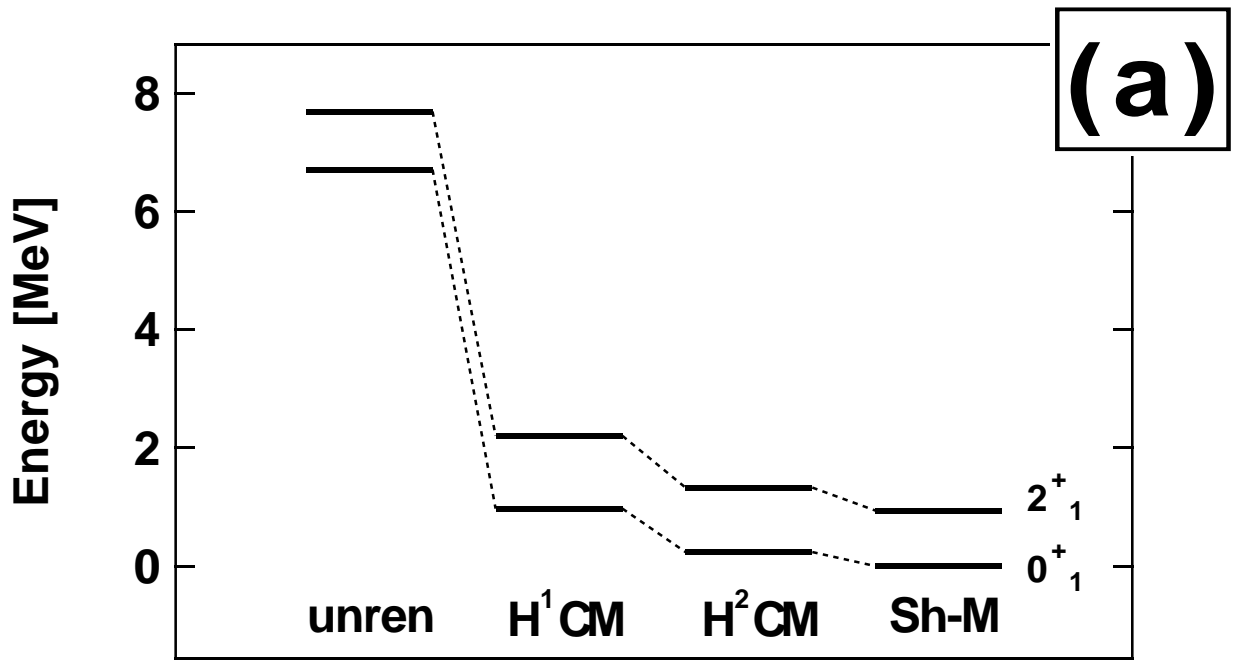
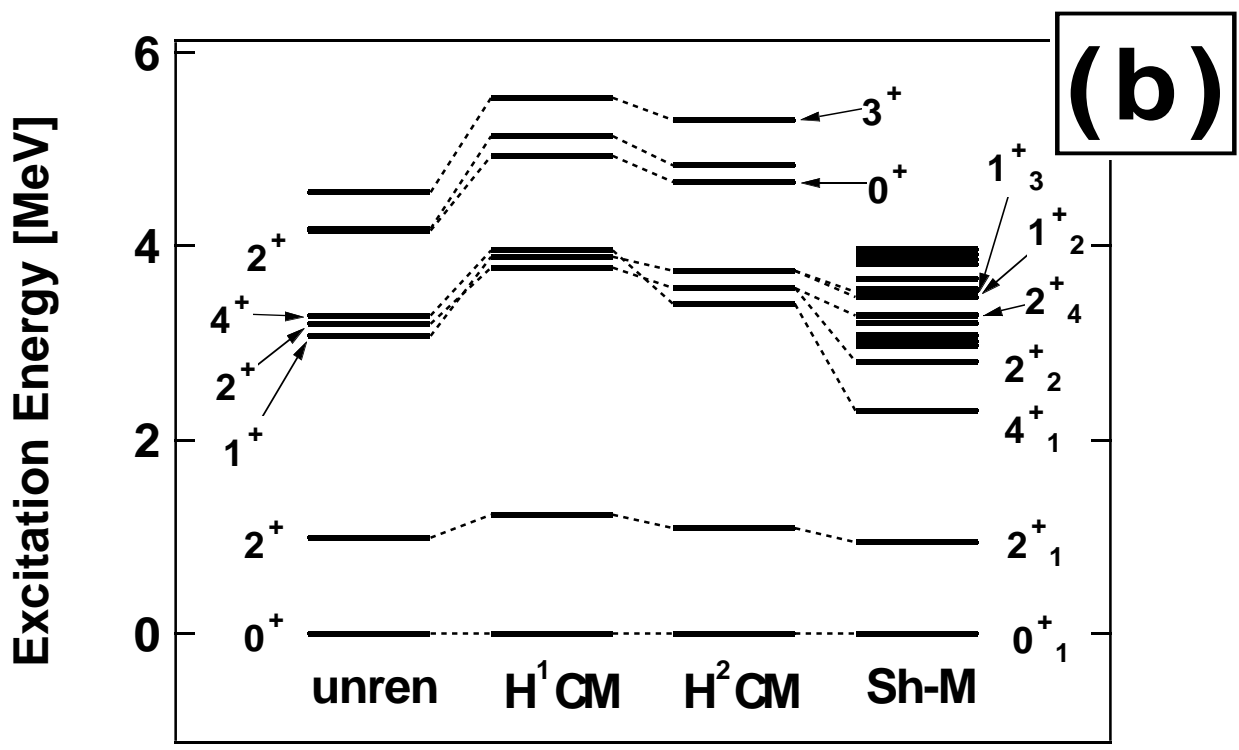
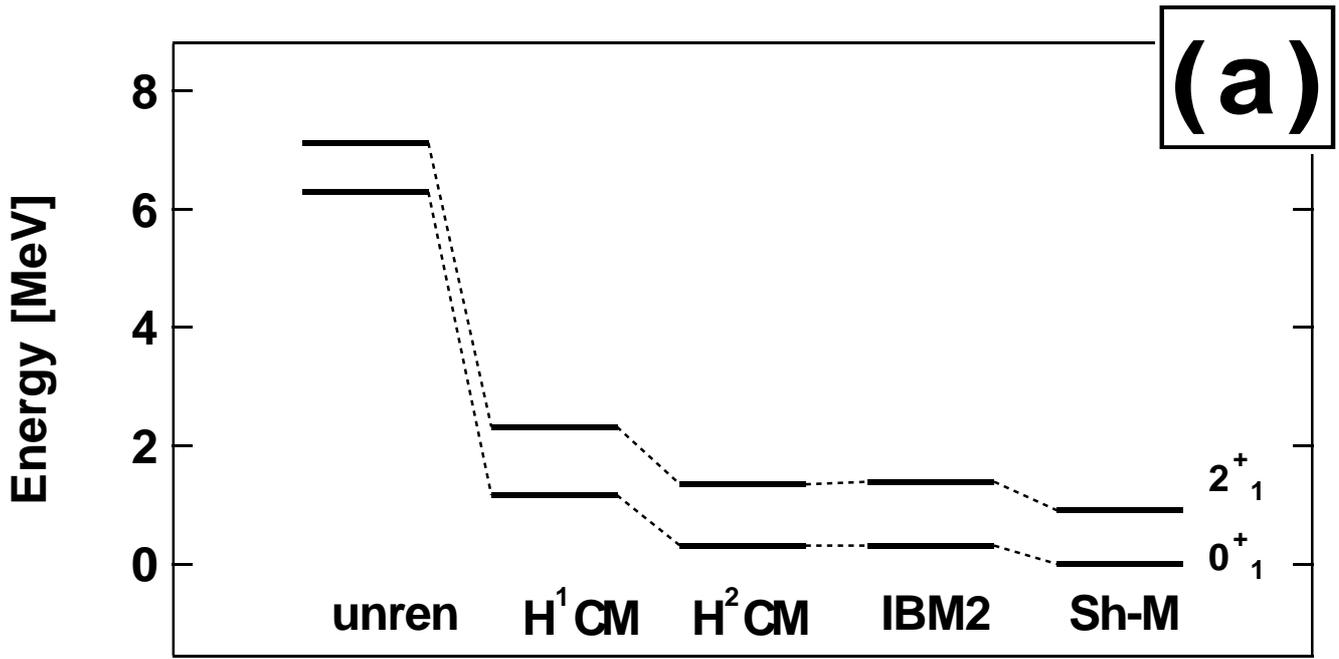
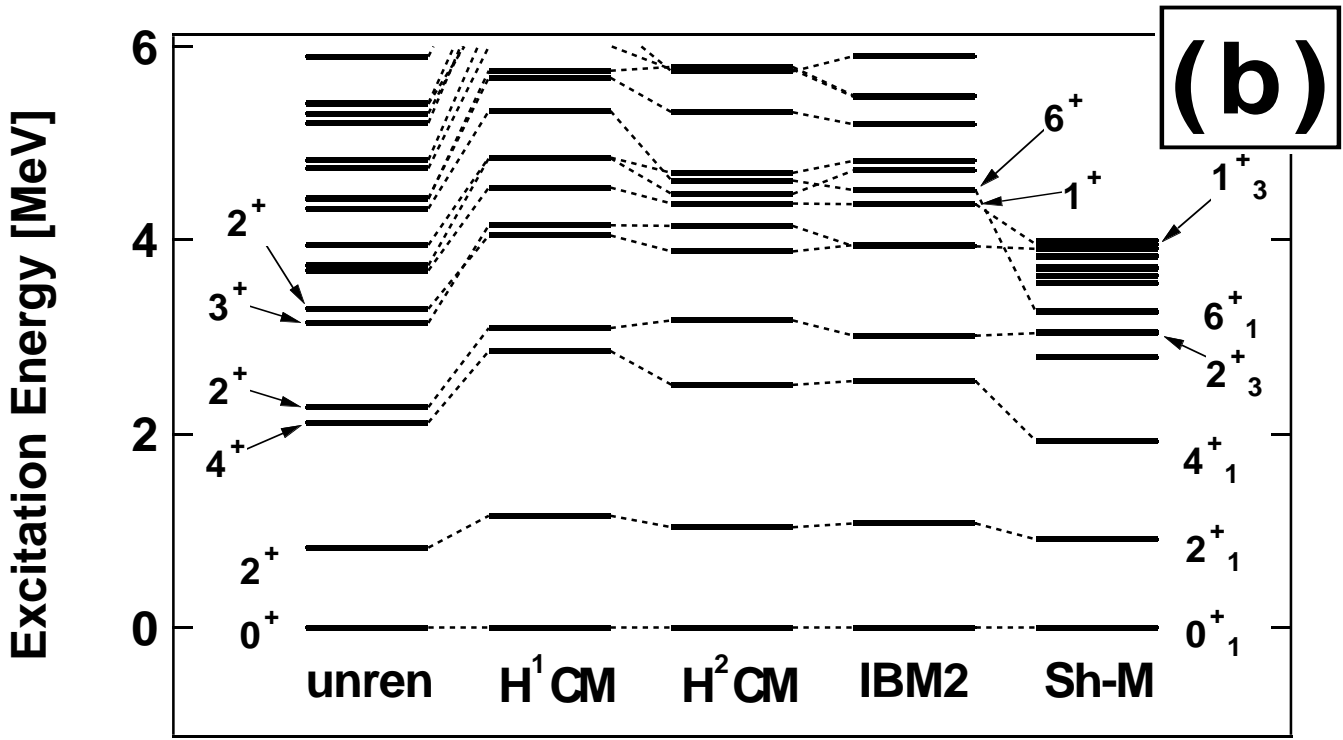
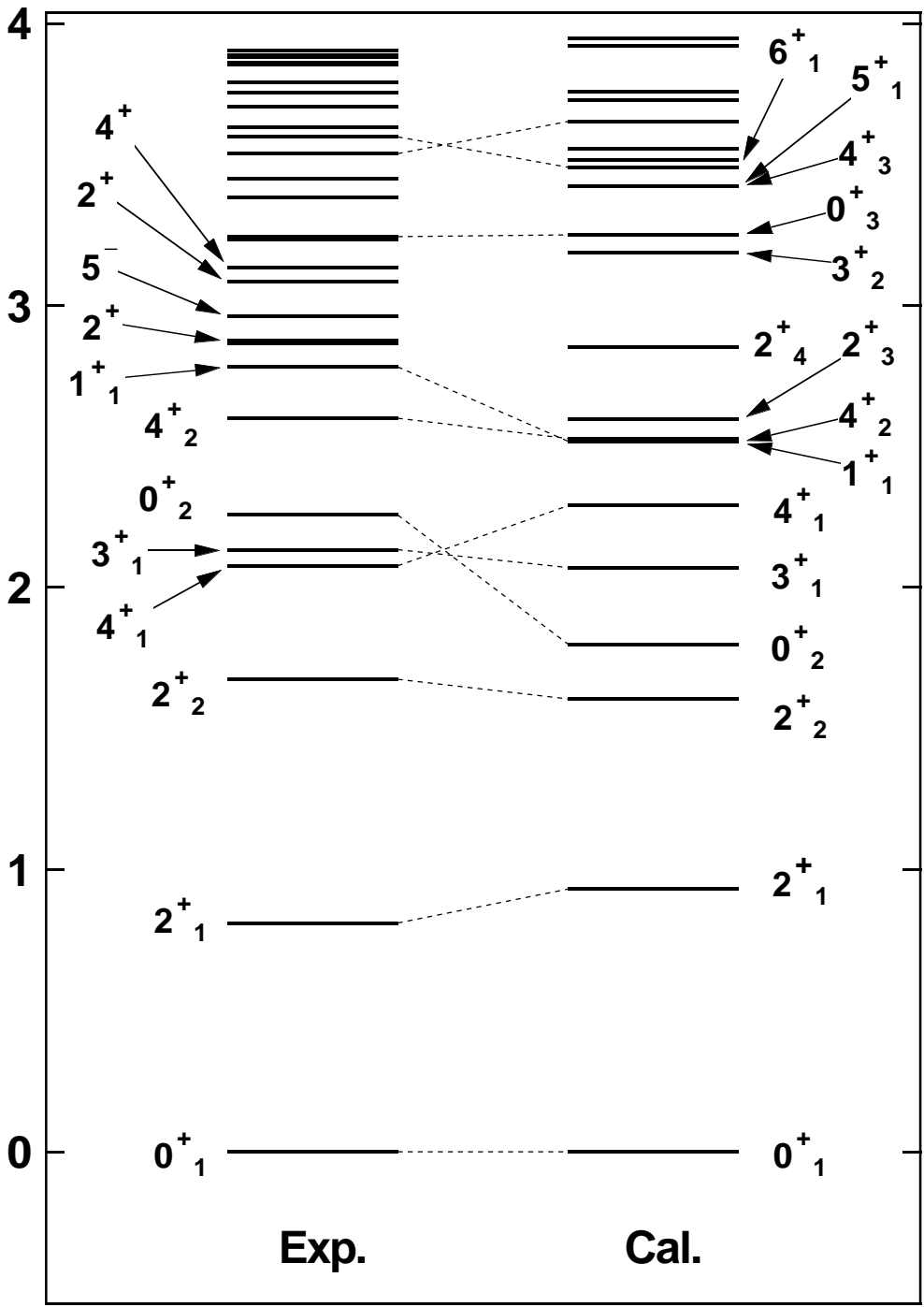
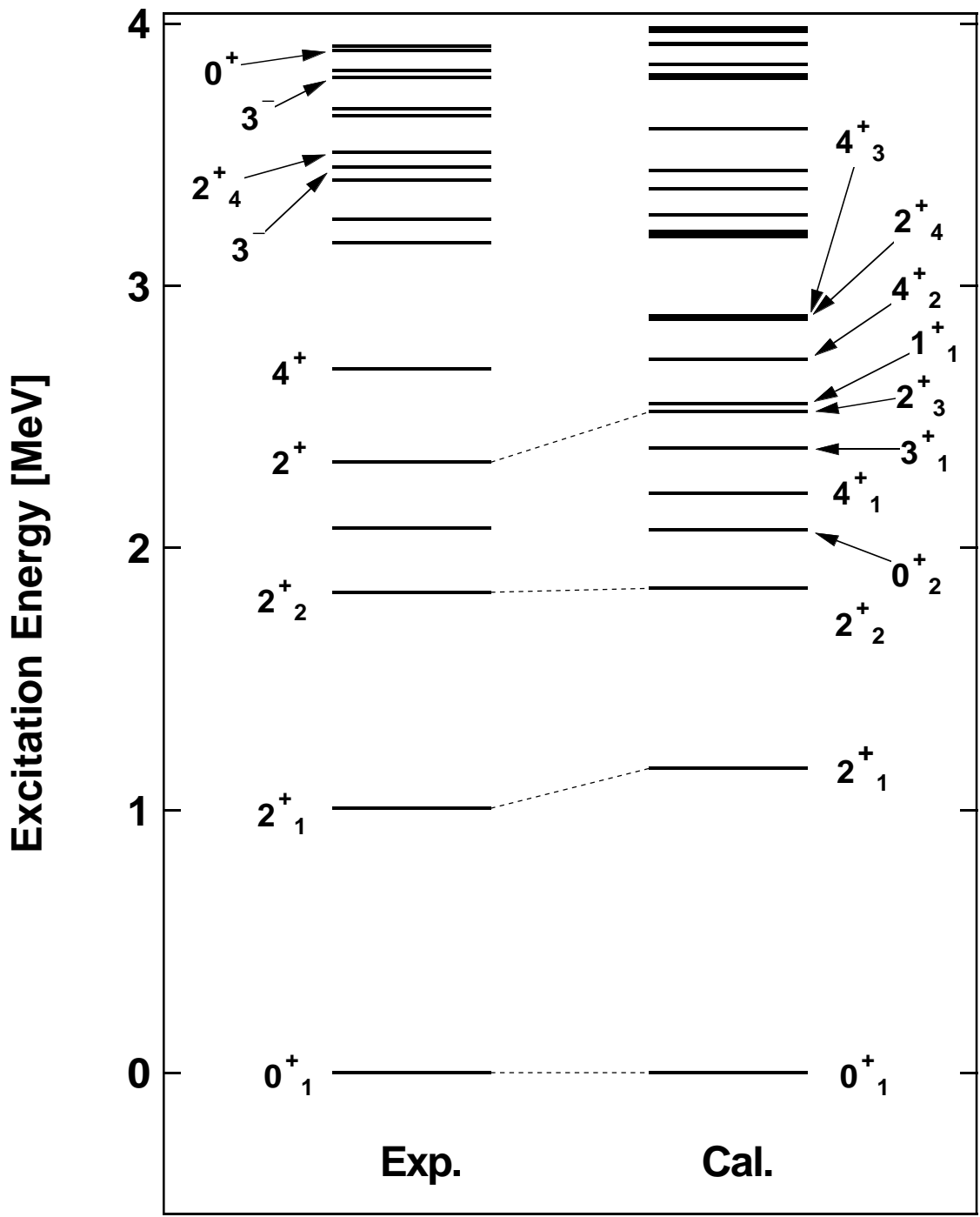
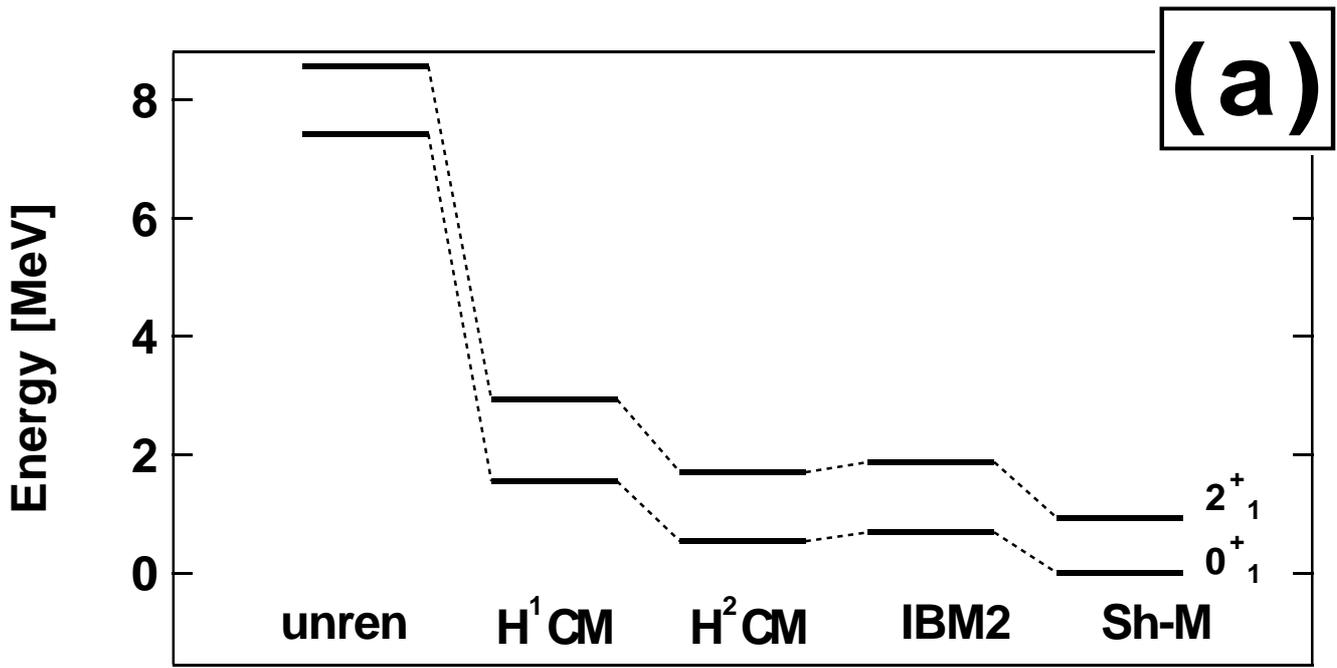
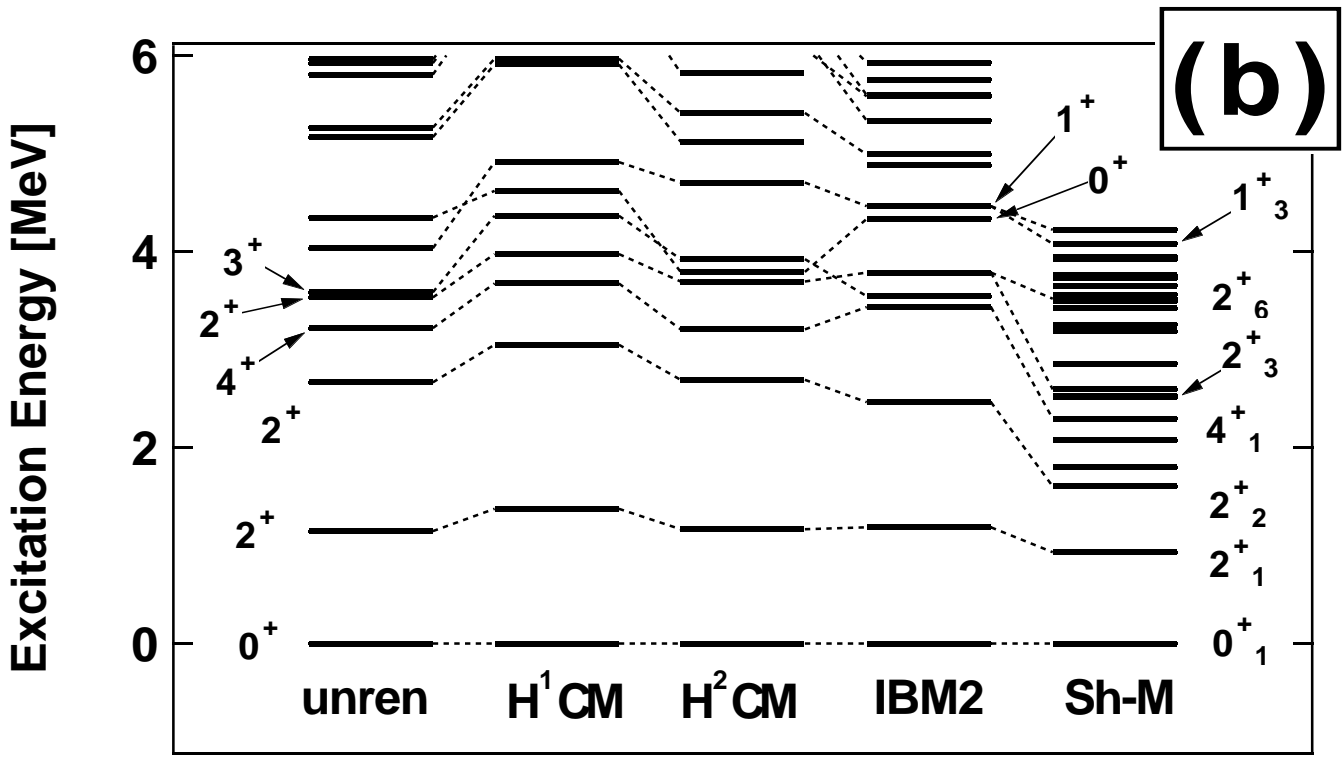
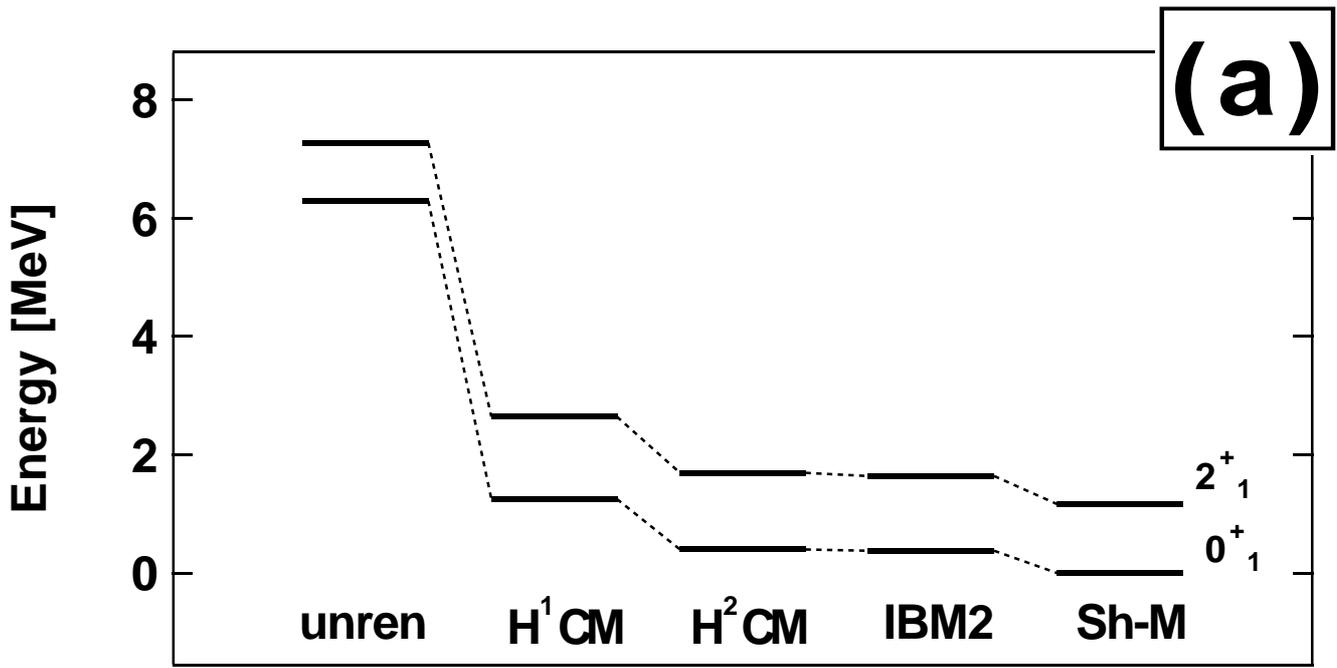
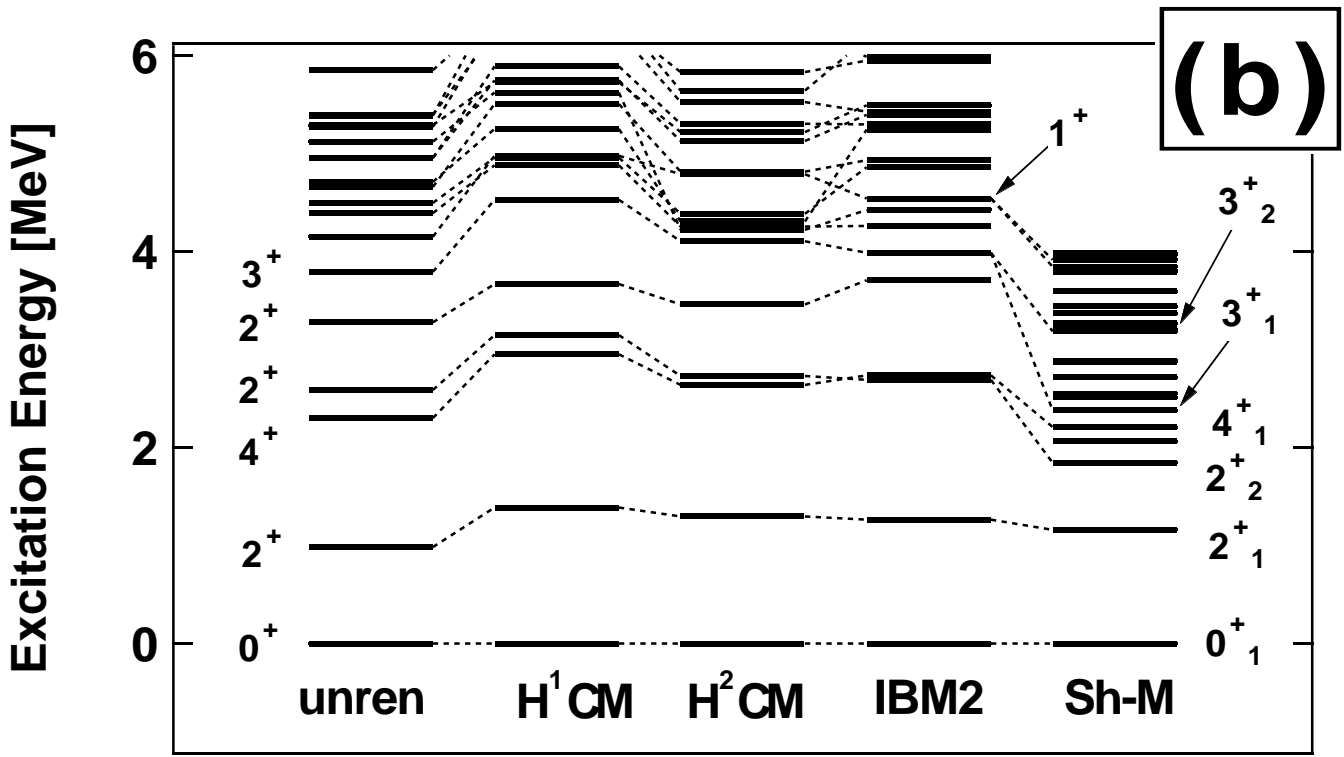
\end{document}